\documentstyle[referee]{mn2e}

\title[NVSS polarization position angles]
{Biases in the polarization position angles in the NVSS point source catalogue}
\author[R.A. Battye {\it et al}]
{R.A. Battye, I.W.A. Browne, N. Jackson\\
Jodrell~Bank~Observatory, University of Manchester, Macclesfield, Cheshire, SK11 9DL U.K.\\}

\begin{document}
\input{psfig}
\label{firstpage}
\maketitle

\begin{abstract}

We have examined the statistics of the polarization position angles determined for point sources in the NRAO-VLA sky survey (NVSS) and find that there is a statistically significant bias toward angles which are multiples of 45 degrees. The formal probability that the polarization angles are drawn from a uniform distribution is exponentially small. When the sample of those NVSS sources with polarizations detected with a signal to noise $\geq$3 is split either around the median polarized flux density or the median fractional polarization, the effect appears to be stronger for the more highly polarized sources. Regions containing strong sources and regions at low galactic latitudes are not responsible for the non-uniform distribution of position angles. We identify CLEAN bias as the probable cause of the dominant effect, coupled with small multiplicative and additive offsets on each of the Stokes parameters. Our findings have implications for the extraction of science, such as information concerning galactic magnetic fields, from large scale polarization surveys.
\end{abstract}


\section{Introduction}

Low frequency measurements of the polarization properties of point sources have been used to deduce information concerning galactic and extragalactic magnetic fields with some success (see, for example, Carilli \& Taylor, 2002; Beck, 2001,  and references therein). Future large scale projects such as the Square Kilometre Array (SKA) intend to make measurements of $\sim 2\times 10^7$ compact polarized extragalactic sources (see discussion in Carilli \& Rawlings, 2004) in order to  understand the nature of the galactic magnetic field structure (Beck \& Gaensler, 2004) and various aspects of cosmic magnetism (Feretti, Burigana \& Ensslin, 2004; Feretti \& Johnston-Hollitt, 2004).

The statistics of polarization position angles (sometimes used in conjunction with structural/jet position angles for spatially resolved sources) have also been used to investigate the global rotation of the universe. In particular, there is a history of claimed detections of global anisotropy and rotation effects using these methods (Birch, 1982; Phinney \& Webster, 1983; Kendall \& Young 1984), the most recent being made by Hutsemekers et al (2005) on the basis of the optical measurements of quasars with magnitudes $\sim 14-15$. However, see Joshi et al (2007) for a discussion of these effects in the context of the radio polarizations measured for flat spectrum sources in the 8.4GHz JVAS/CLASS survey (Jackson et al, 2007).

The veracity of these interesting applications requires the data to be free of systematics to an unprecedented level of accuracy. In any survey of $\sim 2\times 10^7$ sources there is the possibility for systematic effects at the level of $0.02\%$ to have an impact on some statistics computed from the data. The aim of this paper is to investigate the systematics in the measurement of polarization in the NRAO-VLA sky survey (NVSS), the largest survey of point sources with published polarization information presently available (Condon et al, 1998). To our surprise, for such a widely used survey, we have identified an obvious and statistically significant bias in the measured polarization position angles.

Details of the NVSS survey are given in Condon et al (1998). Here we briefly summarize some of the survey details which are relevant to the following discussion. NVSS is a blind continuum survey which covers 82\% of the celestial sphere at a frequency of $\nu=1.4\,{\rm GHz}$ in Stokes parameters $I$, $Q$ and $U$. It was made with the Very Large Array (VLA) operating in its D configuration. A total of $2\times 10^6$ discrete sources were found with flux densities $I>2.5{\rm mJy}$ and the average noise levels were $\sigma_I\approx 0.45\,{\rm mJy}\,{\rm beam}^{-1}$ and  $\sigma_Q=\sigma_U\approx 0.3\,{\rm mJy}\,{\rm beam}^{-1}$. The sky was covered using 217446 overlapping, snapshot images. Each snapshot image was CLEANed (Hogbom 1974) using super-uniform weighting to
optimise the beamshape. In the process of CLEANing large fields, the
effect of CLEAN bias (section~\ref{sec:clean})
can in principle reduce the flux 
density in the final images. In the NVSS survey, the CLEAN procedure
for the total-intensity maps was terminated when the peak residual 
in the residual map fell to 0.75~mJy~beam$^{-1}$. Sources were 
extracted from the total intensity maps using an elliptical 
Gaussian-fitting procedure. In the case of the polarization (Q and U)
images, a small amount of CLEANing was carried out, and the flux
densities were extracted by interpolation at the fitted total-intensity
positions rather than by direct fitting.

In section~\ref{sec:stat} we identify and characterize what we believe is a previously unidentified systematic in the polarization position angles in the published NVSS catalogue. In sections~\ref{sec:clean} and \ref{sec:offset} we discuss the effects of CLEAN bias and offsets on the polarization angle distribution and in section~\ref{sec:data} we show how a combination of these effects can be used to explain the observed biases in the NVSS data. We make an attempt to ``correct'' these biases using a simple prescription in section~\ref{sec:corr}. In the final section we discuss the possible implications of our findings for future observations.

\section{Statistics of polarization position angles in NVSS}
\label{sec:stat}

\begin{figure}
\begin{tabular}{c|c}
{\psfig{figure=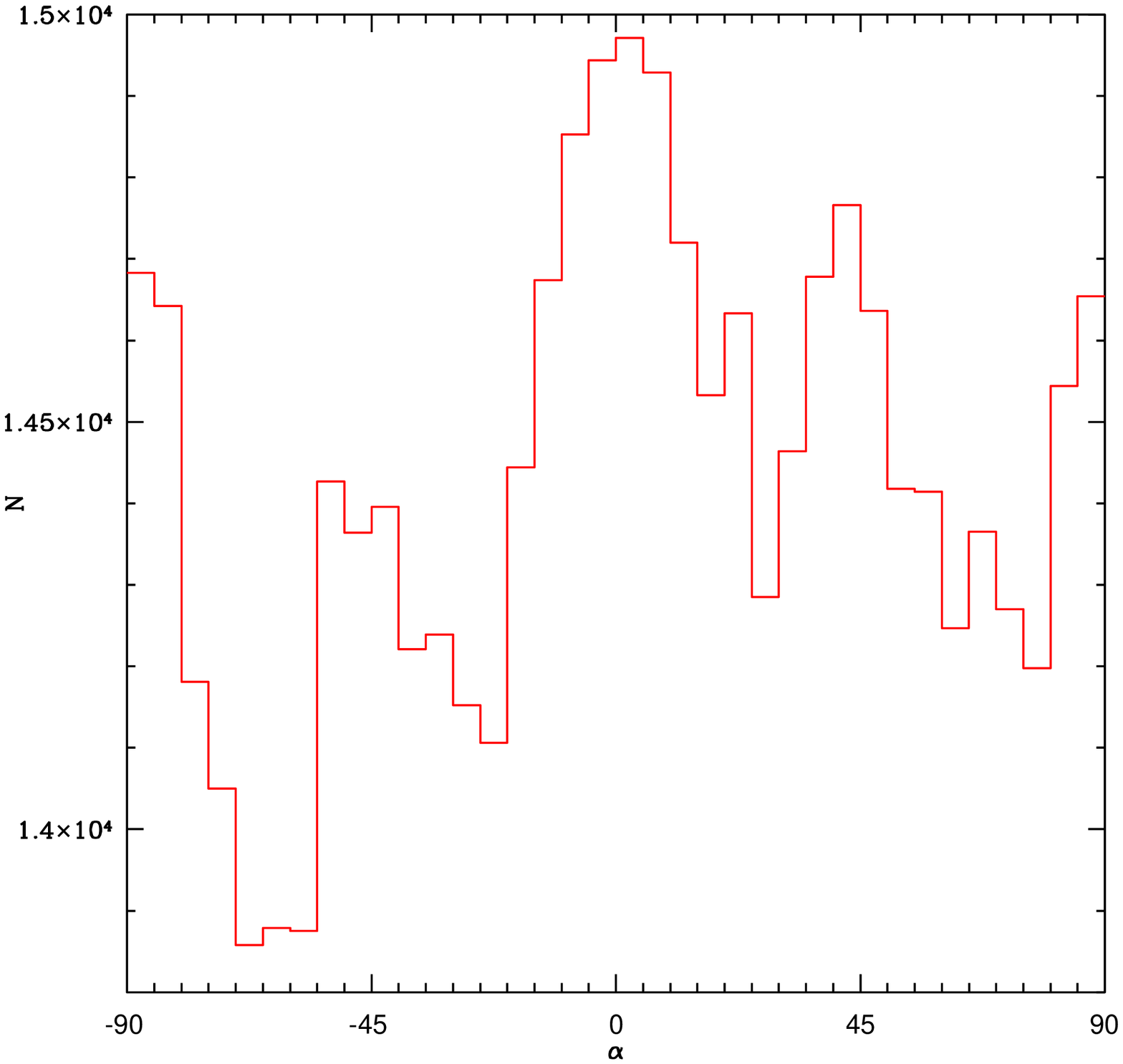,width=8cm}\psfig{figure=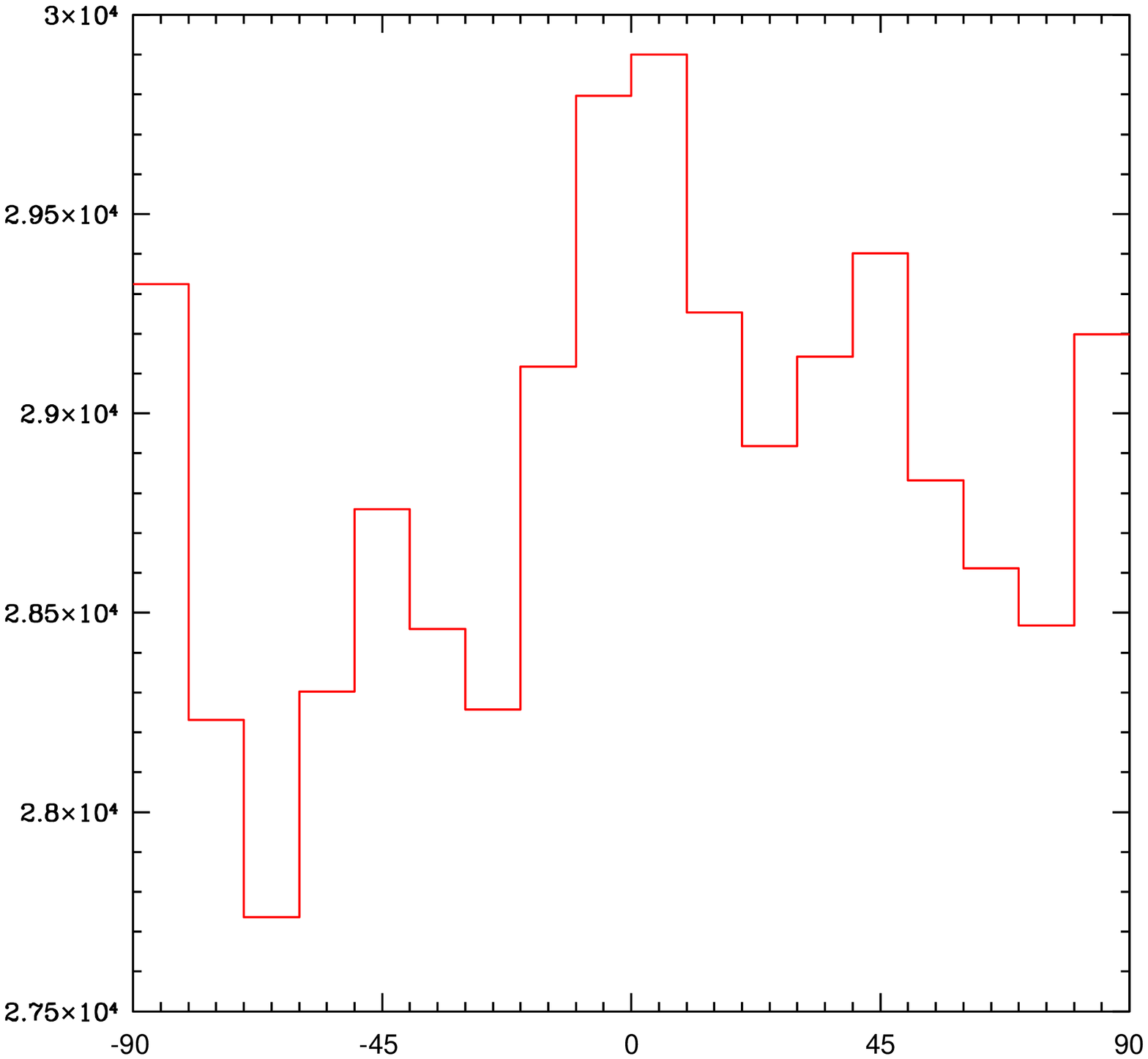,width=8cm}}
\end{tabular}
\begin{tabular}{c|c}
{\psfig{figure=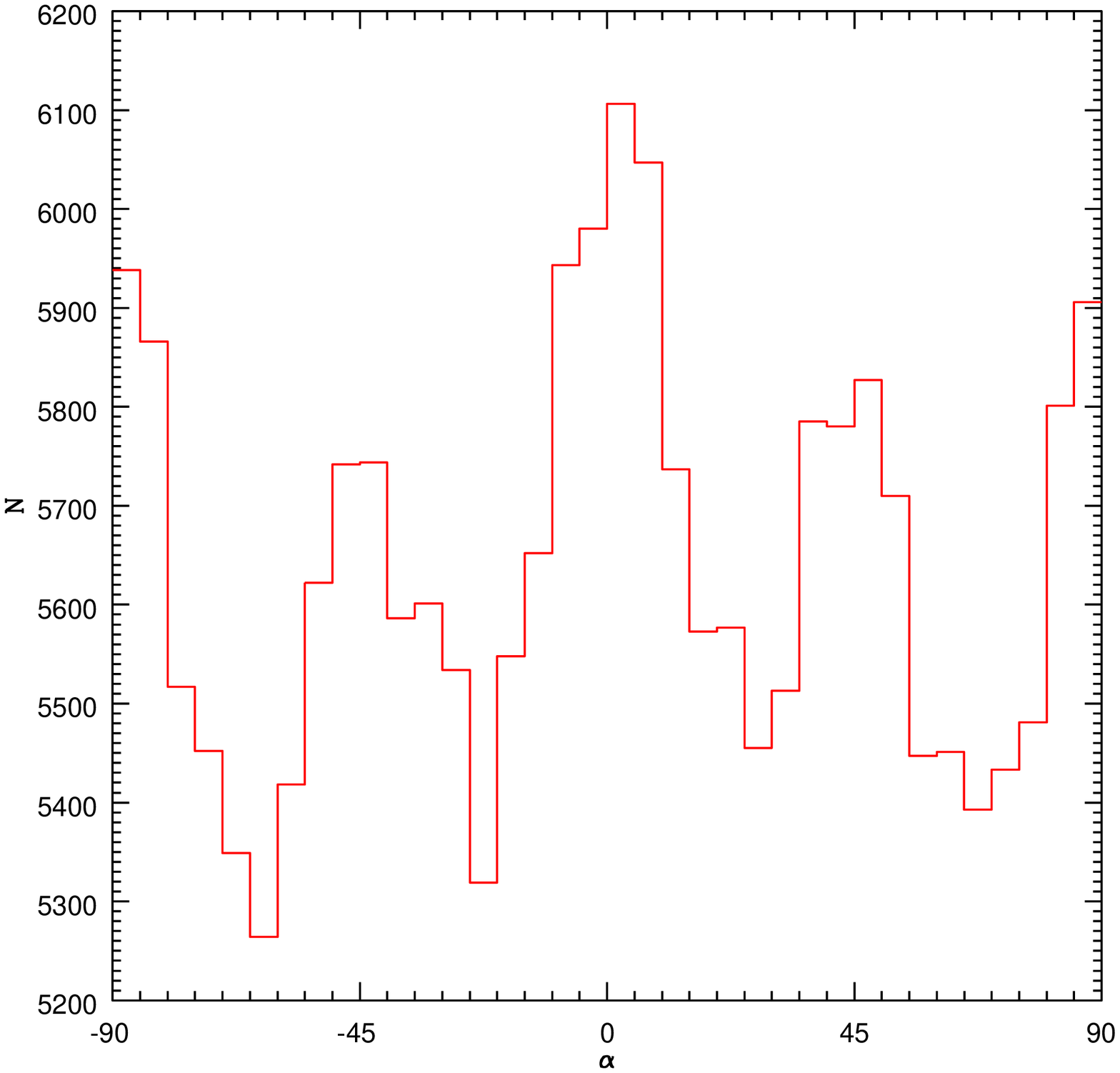,width=8cm}\psfig{figure=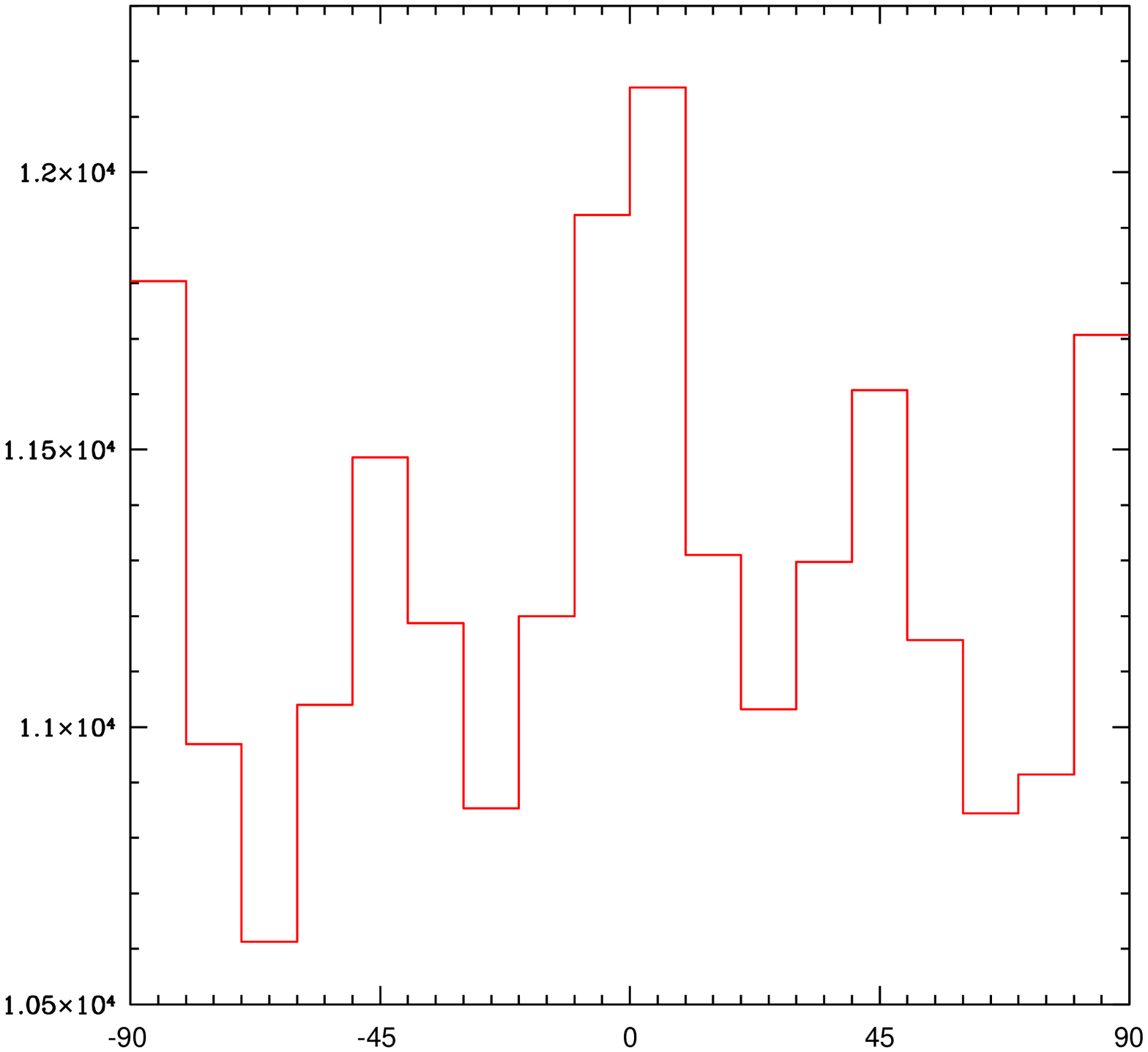,width=8cm}}
\end{tabular}
\caption{Histogram of the polarization position angle for all NVSS sources with flux greater than 10mJy and a quoted angle (top) and those with $P>1{\rm mJy}$ (bottom), using 36 bins (left) and 18 bins (right). Note the apparent bias toward angles which are multiples of $45^{\circ}$ which is present at extremely high significance in each.}
\label{fig:hist}
\end{figure}

We have selected a sample of 519713 sources from the NVSS catalogue with flux $I>10{\rm mJy}$ for which there is a quoted polarization position angle $\alpha$ defined by 
\begin{equation}
\alpha={1\over 2}{\rm tan}^{-1}\left({U\over Q}\right)\,,
\end {equation}
which is in the range $-90^{\circ}\le\alpha\le 90^{\circ}$. We have computed a histogram of these data with bin widths of $5^{\circ}$ and $10^{\circ}$ (36 and 18 bins respectively); the results are presented in Fig.~\ref{fig:hist} (top). It is clear from this that there is a substantive systematic bias toward position angles which are multiples of $45^{\circ}$ and which correspond to states with either pure $Q$ or pure $U$. The amplitude of the bias is around 5-10\% whereas the expected random error on each ($\sim 1\sqrt{N_{\rm bin}}$) is less than $1\%$ in both cases. Since the effect is coherent over a number of bins it is extremely significant. Using a simple $\chi^2$ test one can compute the probability that the observed values of $\alpha$ are selected from a uniform distribution. In both cases the formal probability is $<10^{-20}$. Clearly, the null hypothesis, a uniform distribution, can be excluded at extremely high significance.

The polarized flux is not detected at high signal-to-noise ($S/N$) in a large fraction of the sources and, therefore, we have for much of the subsequent discussion excluded sources with polarized flux density $P<1{\rm mJy}$; the resulting sub-sample contains a total of 203097 sources. Given the noise levels quoted for the survey this corresponds to a sample with a detection of polarization with $S/N> 3$. The equivalent histogram for this sample is presented in Fig.~\ref{fig:hist} (bottom). The bias noted for all sources is still detected at high significance in this sub-sample.  

\begin{figure}
\begin{center}
\hskip 2.0in\psfig{figure=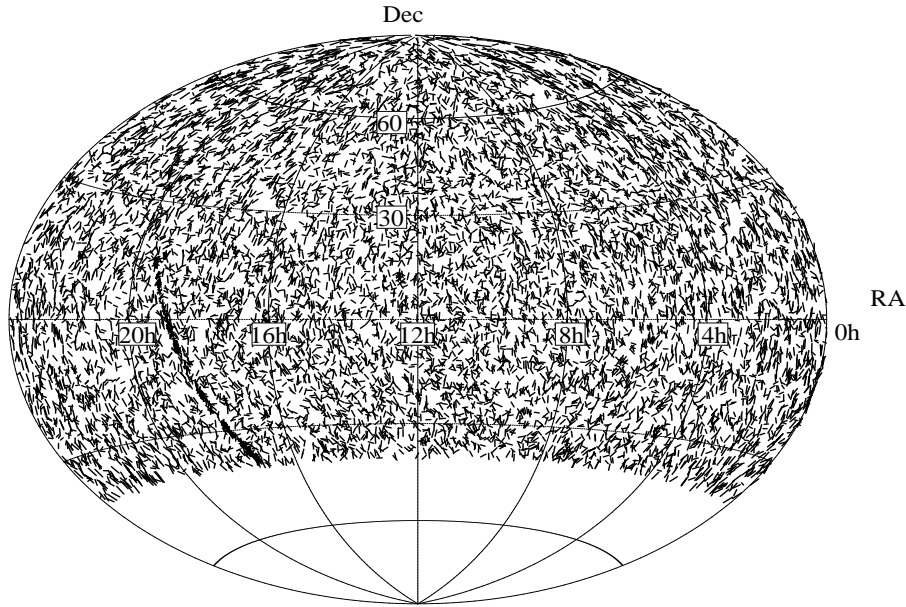,width=12cm,height=8cm}
\end{center}
\caption{The polarization position angles of all NVSS sources with $I>300{\rm mJy}$ using a Hammer-Aitoff projection. Each source is represented by line at the relevant angle to the vertical. The low galactic latitude plane is clearly visible as is the Cygnus region (RA$\sim 20$hr, Dec$\sim 40^{\circ}$).}
\label{fig:aitoff}
\end{figure}

Such a bias should be regarded with some suspicion since a number of possible systematic errors might cause such an effect. Therefore, we have made a number of cuts on the data:
\medskip

\noindent (i) Examination of Fig.~\ref{fig:aitoff} which presents the distribution of polarization angles on the sky for $I>300{\rm mJy}$ illustrates that there are regions, that appear to be connected with the positions of known strong sources (for example, the Cygnus region) and low galactic latitude, where the polarization position angles are strongly correlated. It is conceivable that some of the catalogued sources in regions around strong sources are not actual objects, but are created by the aliasing power from the nearby single strong source via the sidelobes. Moreover, a similar effect might also be present due to strong emission from the galactic plane/centre. We would like to be sure that these sources are not responsible for the bias presented in Fig.~\ref{fig:hist}. In order to investigate this we have made two cuts, the first which should exclude all these regions and the other which is very conservative: (a) we have excluded sources which are within $4^{\circ}$ from sources with flux density $I>8{\rm Jy}$ and also those within $10^{\circ}$ of the galactic plane ($|b|<10$) and (b) exclude regions within $4^{\circ}$ of source with flux density $I>2{\rm Jy}$ and those with $|b|<30^{\circ}$.  The results are almost identical to the case when all sources are included and we conclude that aliasing from strong sources is not responsible for the bias.

\begin{figure}
\begin{tabular}{c|c}
{\psfig{figure=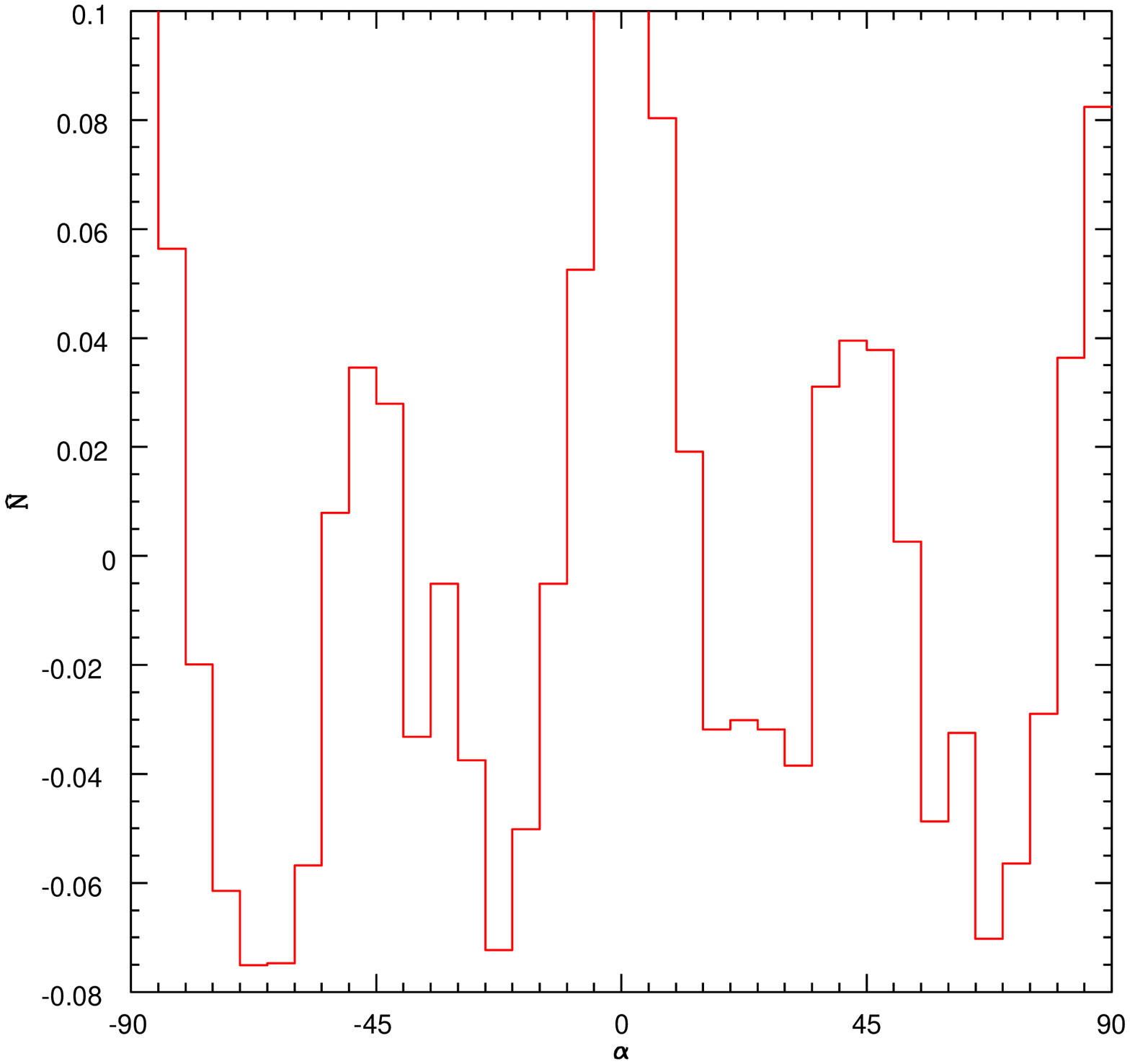,width=8cm}\psfig{figure=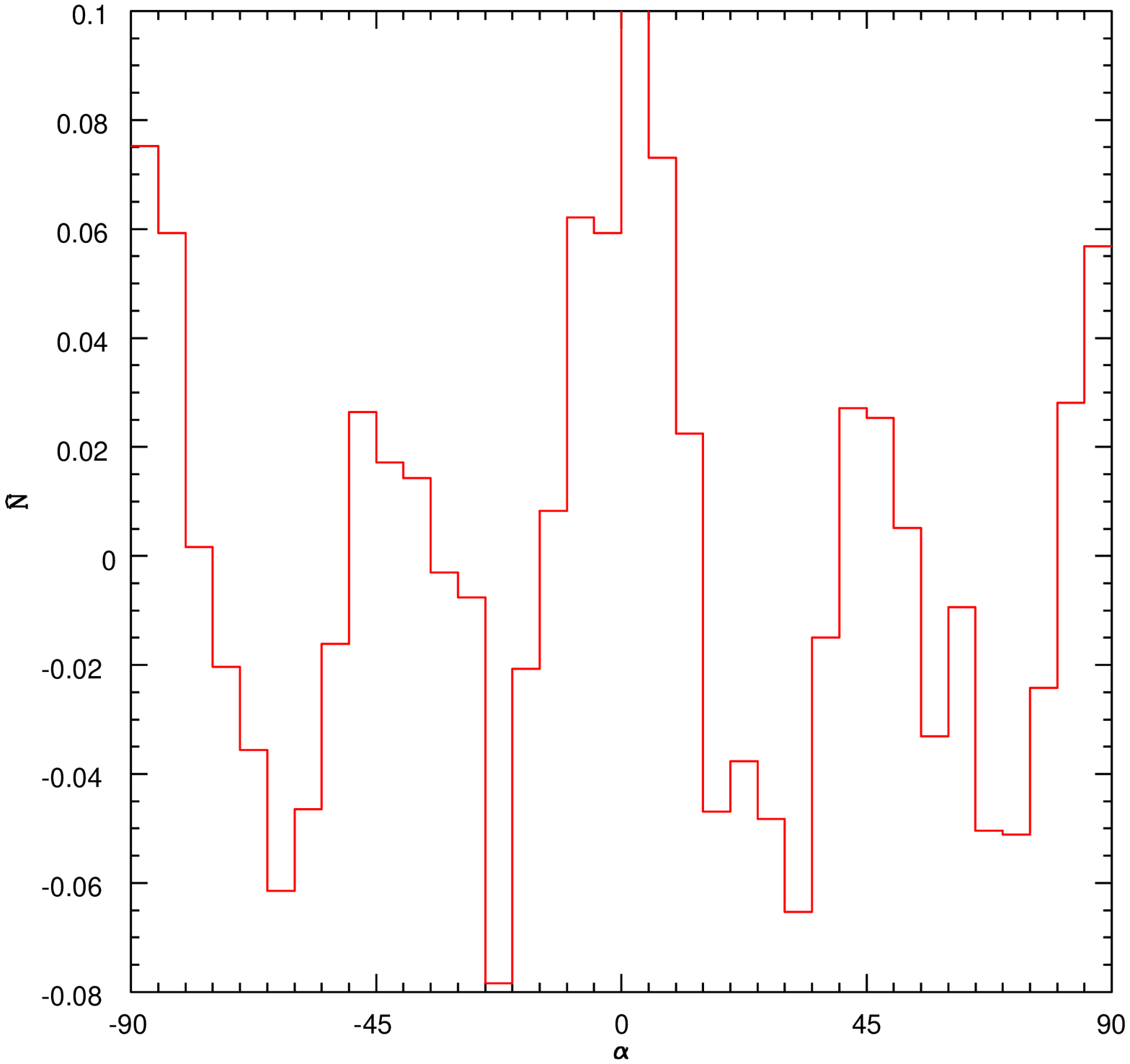,width=8cm}}
\end{tabular}
\begin{tabular}{c|c}
{\psfig{figure=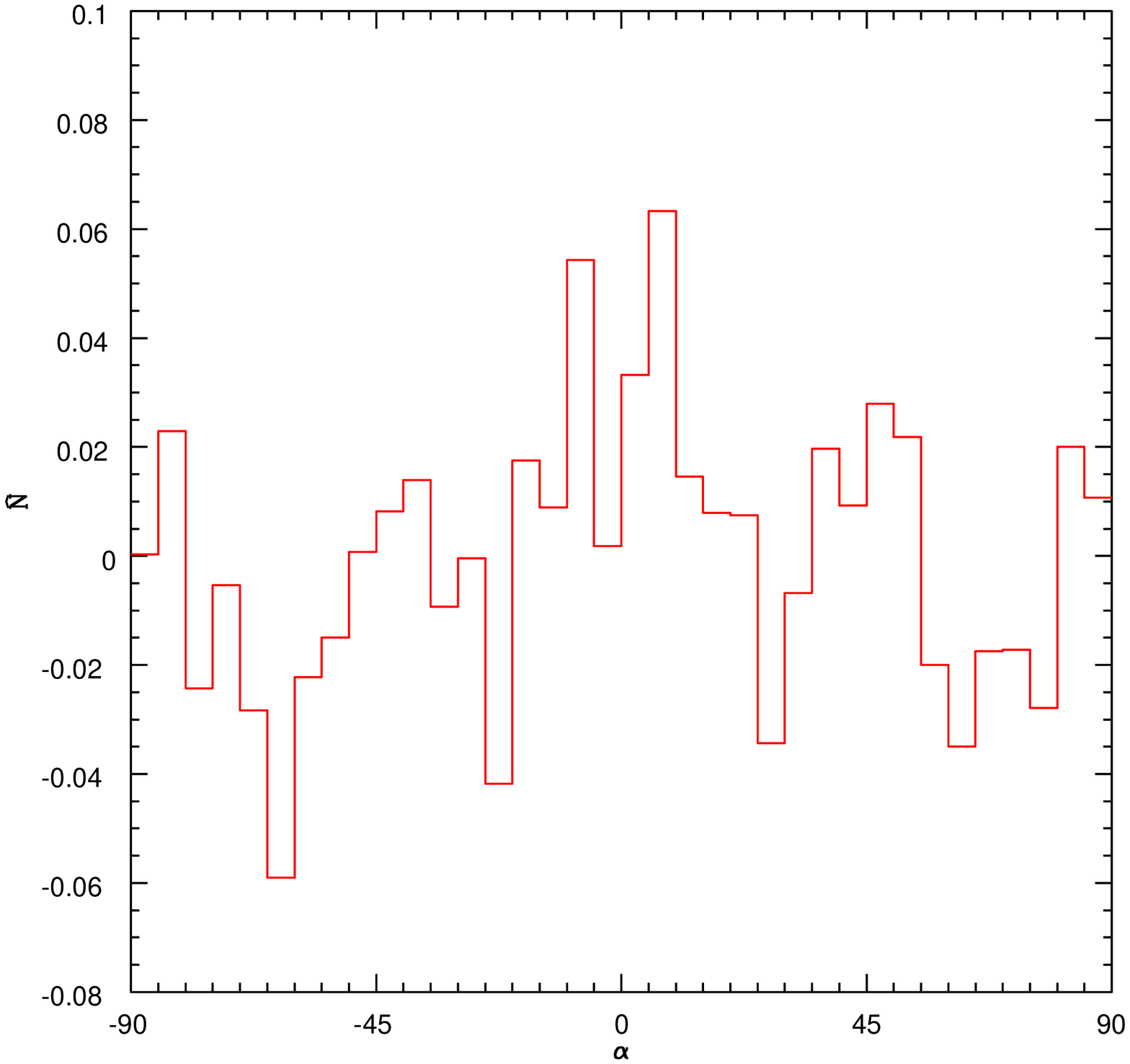,width=8cm}\psfig{figure=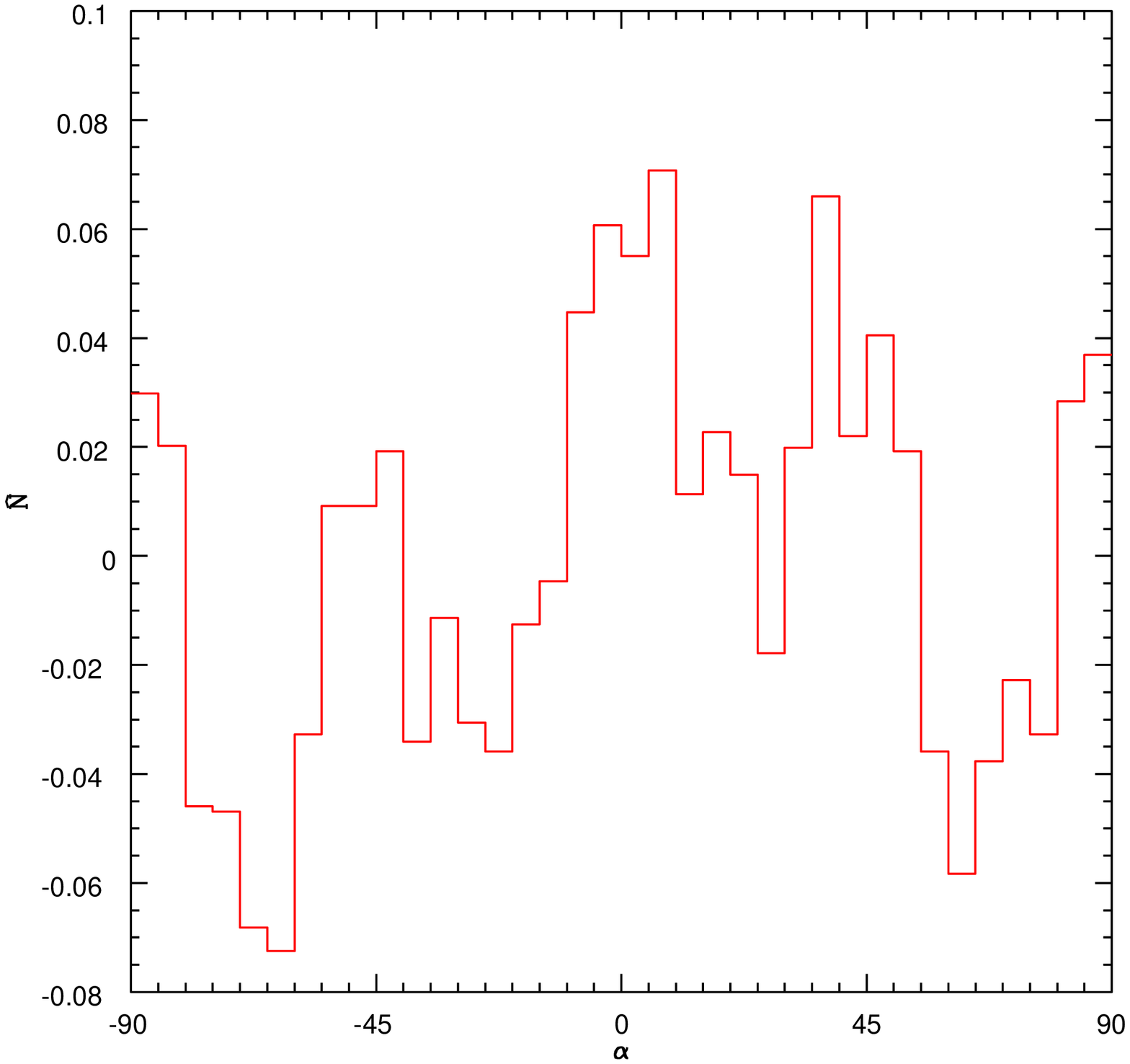,width=8cm}}
\end{tabular}
\caption{Splits of the data illustrating the apparent independence of the effect presented in Fig.~\ref{fig:hist}. (left) the sample split about the median polarized flux $P_{\rm med}\approx 2.1{\rm mJy}$; (right) the sample split about the median polarization fraction ${\Pi}_{\rm med}=0.066$. In both cases the top is the sample above the median and the top that below the median. Note that we have used the fractional difference from the mean, $\hat N$, in these figures as opposed to the number in each bin $N$ which was used in Fig.~\ref{fig:hist}.}
\label{fig:percent}
\end{figure}

\medskip 
\noindent (ii) We have split the sub-sample of sources with polarization $S/N>3$ about the median polarized flux $P_{\rm med}(=\sqrt{Q^2+U^2})\approx 2.1{\rm mJy}$ and about the median polarized fraction ${\Pi}_{\rm med}(=P/I)\approx 0.066$ and the results are presented in Fig.~\ref{fig:percent}. In this figure we have used the fractional differences ${\hat N}=(N-{\bar N})/{\bar N}$ from a constant mean ${\bar N}$ are used instead of the total number $N$, since the samples are of different size. The level of fluctuations in the distribution for $P>P_{\rm med}$ is a little higher than for the whole sample, whereas for $P<P_{\rm med}$ it is more uniform, but still shows some weak evidence for the effect. This is to be expected since the low $P$ sample will be affected by noise, which will have the tendency to make the distribution of $\alpha$ more uniform, even in the presence of a systematic effect. The histogram for $\Pi>\Pi_{\rm med}$ is similar to that for $P>P_{\rm med}$, while that  $\Pi<\Pi_{\rm med}$ still appears to exhibit similar biases, albeit at a lower level. It is remarkable that the bias appears to be strongest in the regime $P>P_{\rm med}$ and $\Pi>\Pi_{\rm med}$ where the measured values of $Q$ and $U$ should be most reliable. 

\medskip
On the basis of the discussion above we conclude that there are biases in the polarization position angles of sources which are in the NVSS catalogue. The bias toward multiples of $45^{\circ}$ suggests some kind of instrumental origin. Moreover, any effect which is present in the sky, such as those discussed in Hutsemekers et al (2005) and Joshi et al (2007), would have been suppressed by Faraday rotation at this relatively very low frequency. In subsequent sections we will attempt to explain these data in terms of various systematic effects.

Since the biases in the polarization angle histograms appear to be periodic, it seems interesting to characterize them using a phenomenological model created from sinusoidal waves with different periodicities. In sections \ref{sec:clean} and \ref{sec:offset} we will argue that there are three separate harmonics involved each of which is a result of a very different kind of systematic that one might expect to be present in the data at some level. In particular we will write the fractional difference from the mean as
\begin{equation}
{\hat N}={N-{\bar N}\over {\bar N}}=A\cos 8\alpha + B\cos 4\alpha + C\cos 2(\alpha-\alpha_0)\,,
\label{phen}
\end{equation}
where $A$, $B$, $C$ and $\alpha_0$ are parameters.

\begin{figure}
\begin{tabular}{c|c}
\psfig{figure=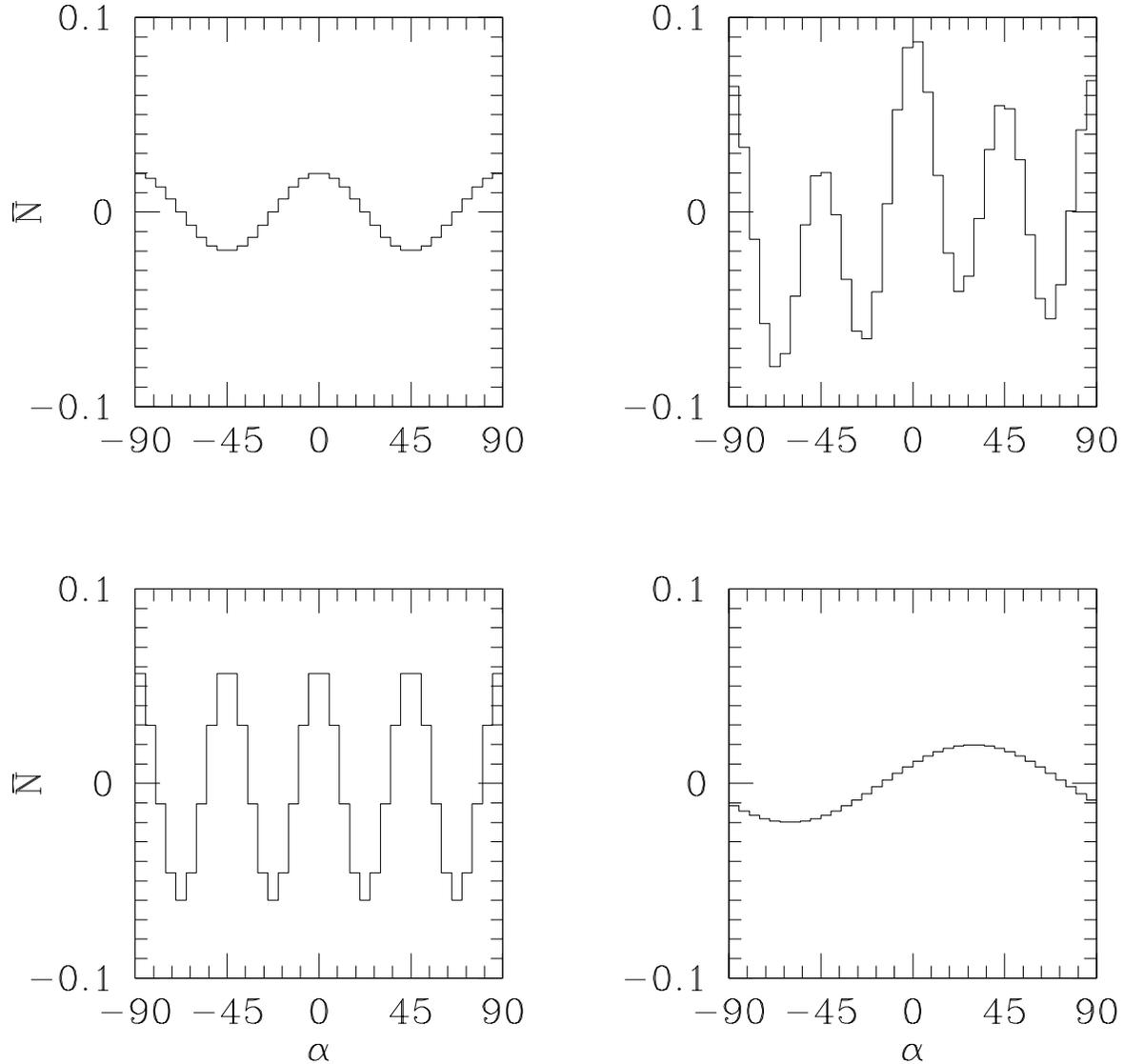,width=16cm,height=16cm}
\end{tabular}
\caption{A simple phenomenological model for the observed biases: (bottom left) ${\hat N}_1=0.06\cos8\alpha$; (top left) ${\hat N}_2=0.02\cos 4\alpha$; (bottom right ${\hat N}_3=0.02\cos 2(\alpha-30^{\circ})$; (top right) the combination of all three ${\hat N}={\hat N}_1+{\hat N}_2+{\hat N}_3$.}
\label{fig:char}
\end{figure}

The main effect in the data has period $45^{\circ}$ corresponding to the maxima at $-90^{\circ}$, $-45^{\circ}$, $0^{\circ}$, $45^{\circ}$ and $90^{\circ}$ in Fig.~\ref{fig:hist} which is represented by the $\cos 8\alpha$ term in (\ref{phen}). In addition to this there appears to be a modulation of the peak heights such that those at $-90^{\circ}$, $0^{\circ}$ and $90^{\circ}$ are higher than those at $-45^{\circ}$ and $45^{\circ}$. This is represented by the $\cos 4\alpha$ term. Both these harmonics seem to be present both in the whole sample and in the sample with just $P>1{\rm mJy}$ in Fig.~\ref{fig:hist}. There appears to be an additional modulation in the whole sample, which is less obvious that with $P>1{\rm mJy}$, suggesting that this effect is only important for the sources with low polarized flux. It appears that there is a modulation  with period $180^{\circ}$ centred around $\alpha=30^{\circ}$ which increases the peak at $45^{\circ}$ relative to that at $-45^{\circ}$ which is represented by the $\cos 2(\alpha-\alpha_0)$ term. We have illustrated these three effects and their combination in Fig.~\ref{fig:char} for $A=0.06$, $B=0.02$, $C=0.02$ and $\alpha_0=30^{\circ}$; parameters which were decided by trial-and-error. It is clear that their combination looks very similar to the histogram presented in Fig.~\ref{fig:hist}.

\section{Systematic effects on polarization measurements}

The model presented in the previous section is purely phenomenological. In this section we consider possible real systematic effects which could lead to the biases observed in the data. Remarkably, there are reasonable effects which would lead to each of the terms in (\ref{phen}).

\subsection{CLEAN bias}
\label{sec:clean}

CLEAN bias is a known phenomenon related to use of the CLEAN algorithm (Hogbom 1974) on radio synthesis data. Its effect on the data from the NVSS, and an explanation for its working, is discussed in some detail by Condon et al.(1998). Becker et al. (1995) also discuss its effect on the FIRST data.

CLEAN bias produces a reduction in the central flux density of sources by scattering flux from the source over a wide area. This reduction becomes progressively more severe as more CLEAN cycles are used. In the case of the NVSS survey, Condon et al. (1998) state that the average clean bias in the NVSS images is approximately $-0.3$~mJy~beam$^{-1}$. Such a bias is relatively benign in the case of total intensity, since it just leads to a small bias in the peak flux density, which is only important for weak sources and can be corrected for.

Its effects on Stokes $Q$ and $U$ maps is much more significant because of (i) the generally low levels of polarization in radio sources and (ii) $Q$ and $U$ can be both positive and  negative. The CLEAN bias effect, which  is to reduce the magnitude toward zero, can have a significant effect on distribution of $\alpha$ even for very small amounts of CLEANing. Since it acts proportionately more on either $Q$ or $U$, which ever is lower. Less information is provided on the CLEANing of the $Q$ and $U$ data by the NVSS team; they state that the images are ``lightly cleaned'' since each of the images contain only a few sources (Condon et al. 1998), suggesting also that no correction is made for clean bias.

We  model the effects of CLEAN bias by assuming that the observed Stokes parameters $(Q_{\rm obs}, U_{\rm obs})$ are related to their true values $(Q_{\rm true},U_{\rm true})$ by the addition or subtraction of $\epsilon_{\rm C}(>0)$ so as to reduce the measured value toward zero. That is, if $Q_{\rm true}>\epsilon_{\rm C}$ then $Q_{\rm obs}=Q_{\rm true}-\epsilon_{\rm C}$, whereas if $Q_{\rm true}<-\epsilon_{\rm C}$ then $Q_{\rm obs}=Q_{\rm true}+\epsilon_{\rm C}$. If $-\epsilon_{\rm C}<Q_{\rm true}<\epsilon_{\rm C}$ then $Q_{\rm obs}=0$. An equivalent procedure is also used  for $U$. The value of $\epsilon_{\rm C}$ can be related to the number of times CLEAN subtracts a component from the measured values of $Q$ and $U$, $N_{\rm C}$. Using simulations with $\sigma_{Q}=\sigma_{U}=300\mu{\rm Jy}$ we have deduced that $\epsilon_{\rm C}\approx 150\mu{\rm Jy}\log_{10}(N_{\rm C})$; it is likely that the coefficient is a function of the noise level. 

\begin{figure}
\begin{tabular}{c|c}
\hskip -1.3cm \psfig{figure=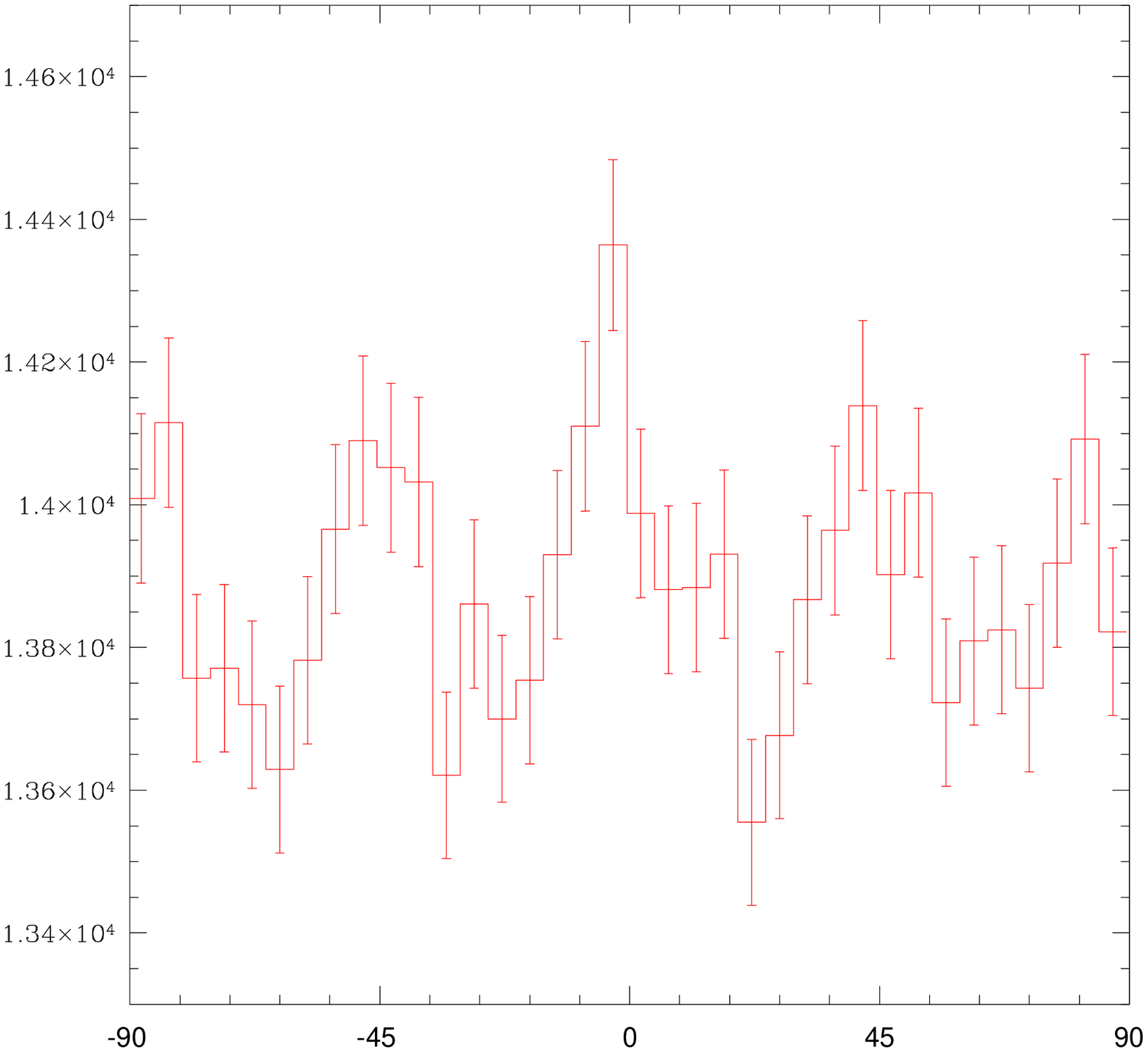,width=9cm,height=9cm}\psfig{figure=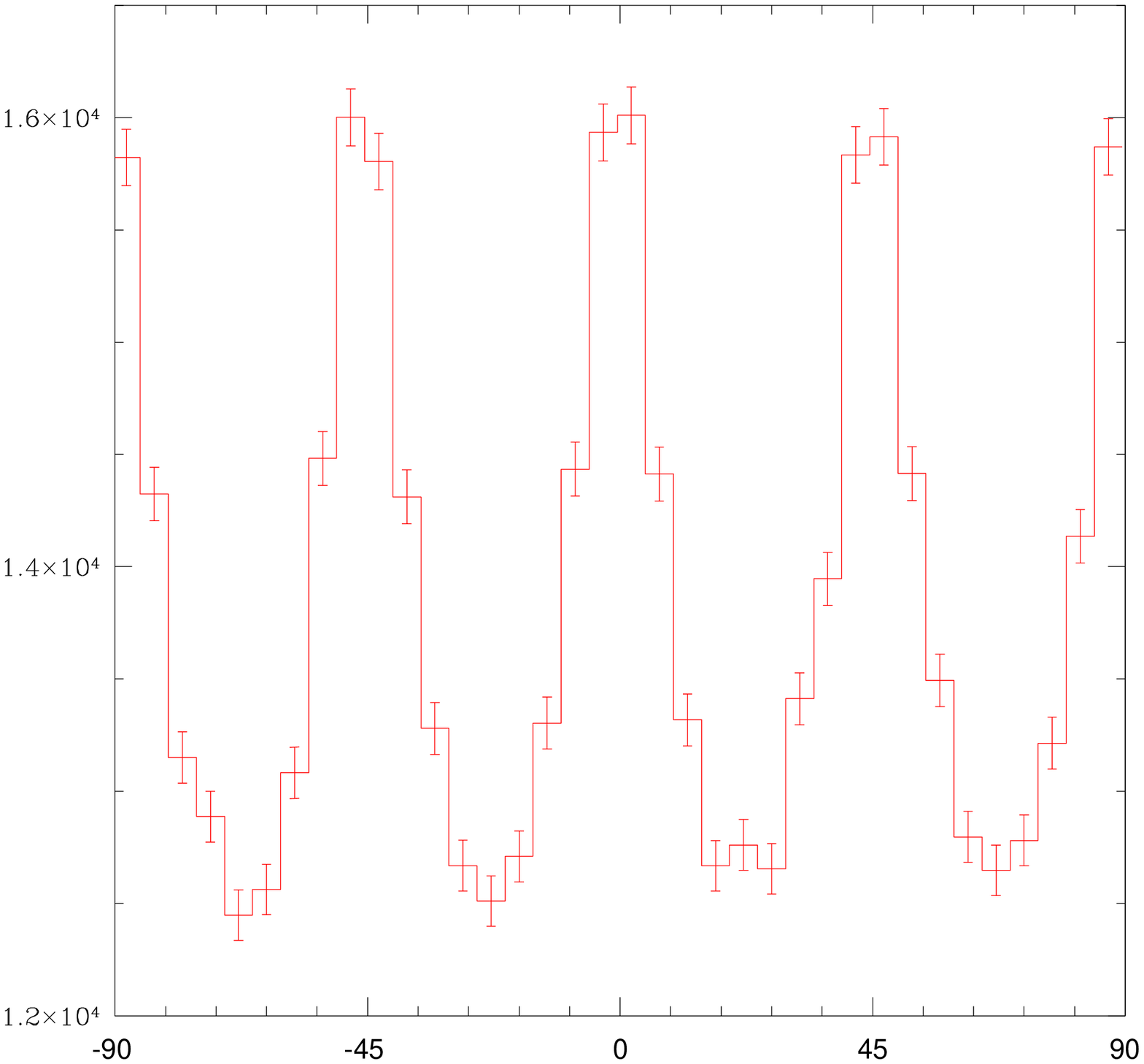,width=9cm,height=9cm}
\end{tabular}
\caption{Histogram of $\alpha$ for a simulated NVSS-like survey containing 500000 sources with the effects of CLEAN bias included. (left) $\epsilon_{\rm C}=15\mu{\rm Jy}$ and (right) $\epsilon_{\rm C}=300\mu{\rm Jy}$. In both cases there is a significant bias toward values of $\alpha$ which are multiple of $45^{\circ}$ as is seen in the histogram of the observed data. The histogram can be modelled by ${\hat N}\propto\cos 8\alpha$. The errorbars quantify the random Poisson error on each bin.}
\label{fig:cleaneffect}
\end{figure}

In order to investigate the effects of CLEAN bias on the histogram of $\alpha$, we have produced simulated catalogue of sources with similar properties to that of the NVSS. In particular we have produced a sample of 500000 sources with polarized fluxes between $0.4{\rm mJy}$ and $100{\rm mJy}$ drawn  from a distribution with $dN/dP\propto P^{-1}$, which have polarization angles with a uniform distribution. The true values of the Stokes parameters have the effects of CLEAN bias included as described above and have Gaussian random noise of $0.3{\rm mJy}\,{\rm beam}^{-1}$ added. The resulting distribution of position angles is presented in Fig.~\ref{fig:cleaneffect} for $\epsilon_{\rm C}=15\mu{\rm Jy}$ and $300\mu{\rm Jy}$ which have been chosen to illustrate the effect. Both are non-uniform with peaks at multiples of $45^{\circ}$, and that with $\epsilon_{\rm C}=15\mu{\rm Jy}$ appears to have an amplitude capable of explaining the one of the observed effects ($\propto\cos 8\alpha$) in the NVSS data. Effects of this kind are to be expected since we have already pointed out that CLEAN biases both $Q$ and $U$ toward zero, which correspond to multiples of $45^{\circ}$ in polarization position angle.

\subsection{Multiplicative and Additive offsets}
\label{sec:offset}

Although the bias toward multiples of $45^{\circ}$ is the most significant effect in the observed data, we have also identified two other possible effects which are $\propto\cos 4\alpha$ and $\propto\cos 2(\alpha-\alpha_0)$ respectively. In this section we will show that these effects can be modelled by small  multiplicative and additive offsets in the data. In particular we will assume that, ignoring for the moment CLEAN bias,
\begin{equation}
Q_{\rm obs}=(1+m_Q)Q_{\rm true}+a_Q\,,\qquad  U_{\rm obs}=(1+m_U)U_{\rm true}+a_{U}\,,
\end{equation}
that is, the multiplicative offsets are parameterized by $m_{Q}$ and $m_{U}$, and the additive offsets by $a_{Q}$ and $a_{U}$.

Errors in polarization calibration are associated with the determination of
instrumental polarizations for the individual telescopes. The effect of a
slight miscalibration typically appears as a rotation of points in the $(Q,U)$
plane around a fixed point $(Q_0,U_0)$. For a large survey with a large number
of observing epochs, the effect will be a combination of many such offset
rotations. For simplicity, we have modelled this complex effect as a
combination of multiplication and addition in the $(Q,U)$ plane as described
above.

One can compute the systematic error in $\alpha$ due to these two effects to first order
\begin{equation}
\delta\alpha={1\over 4}(m_U-m_Q)\sin 4\alpha-{\sqrt{a_Q^2+a_U^2}\over 2P}\sin 2(\alpha-\alpha_0)\,,
\label{delalpha}
\end{equation}
where $\alpha$ is the true polarization angle, $P$ is the polarized flux and $\tan 2\alpha_0=a_{U}/a_Q$. Converting this expression into the corresponding effect on the histogram requires some thought. Let us consider which direction the value of $\alpha$ is offset due to the first term which represents the multiplicative offset. There are fixed points at $-90^{\circ}$, $-45^{\circ}$, $0^{\circ}$, $45^{\circ}$, $90^{\circ}$. If $m_U-m_Q>0$ then $-45^{\circ}$ and $45^{\circ}$ are attractive with the others repulsive. The opposite is true for $m_U-m_Q<0$. As a general rule we find that each of the trigonometric functions in (\ref{delalpha}) need to have their phase modified by $90^{\circ}$, that is, 
\begin{eqnarray}
{\hat N}&=&{1\over 4}(m_U-m_Q)\sin (4\alpha-90^{\circ})-{\sqrt{a_Q^2+a_U^2}\over 2P}\sin [2(\alpha-\alpha_0)-90^{\circ}]\cr
&=&{1\over 4}(m_Q-m_U)\cos 4\alpha+{\sqrt{a_Q^2+a_U^2}\over 2P}\cos 2(\alpha-\alpha_0)\,.\label{alphahist}
\end{eqnarray}
 From this we see that the multiplicative offset can give rise to an effect $\propto\cos 4\alpha$ and the additive offset can give rise to one $\propto\cos 2(\alpha-\alpha_0)$. Moreover, we see that the effects due to the additive offsets are suppressed by $1/P$ and hence they would not show up for large $P$.

We note that in the case of the multiplicative offset it is really the ratio $R=(1+m_U)/(1+m_Q)$ which is relevant and it is this which we will use in the subsequent fitting. However, in deriving (\ref{delalpha}) we have made the assumption that $m_U$ and $m_Q$ are small in which case $R\approx 1+m_U-m_Q$. 

\begin{figure}
\begin{tabular}{c|c}
\hskip -1.3cm \psfig{figure=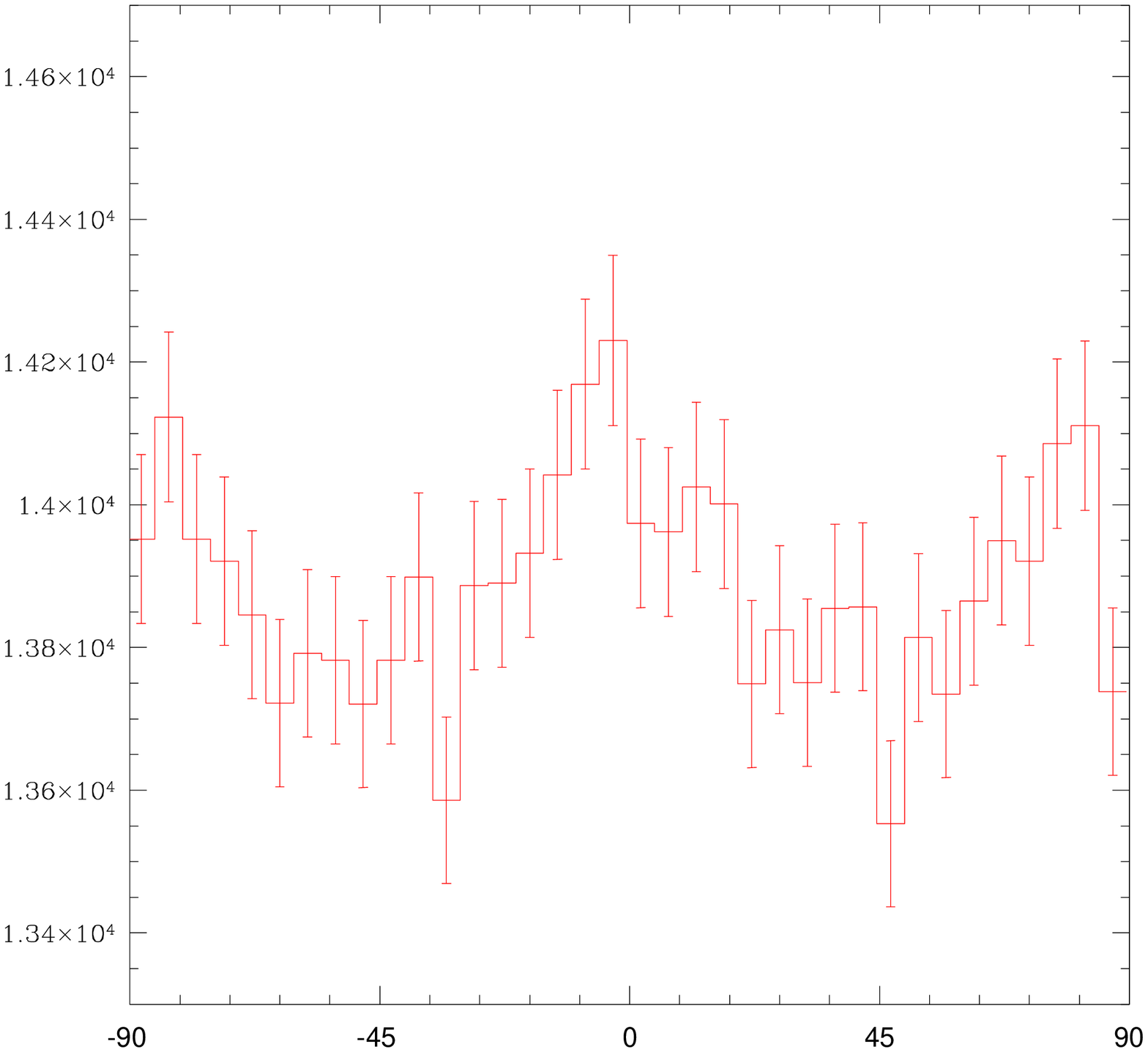,width=9cm,height=9cm}\psfig{figure=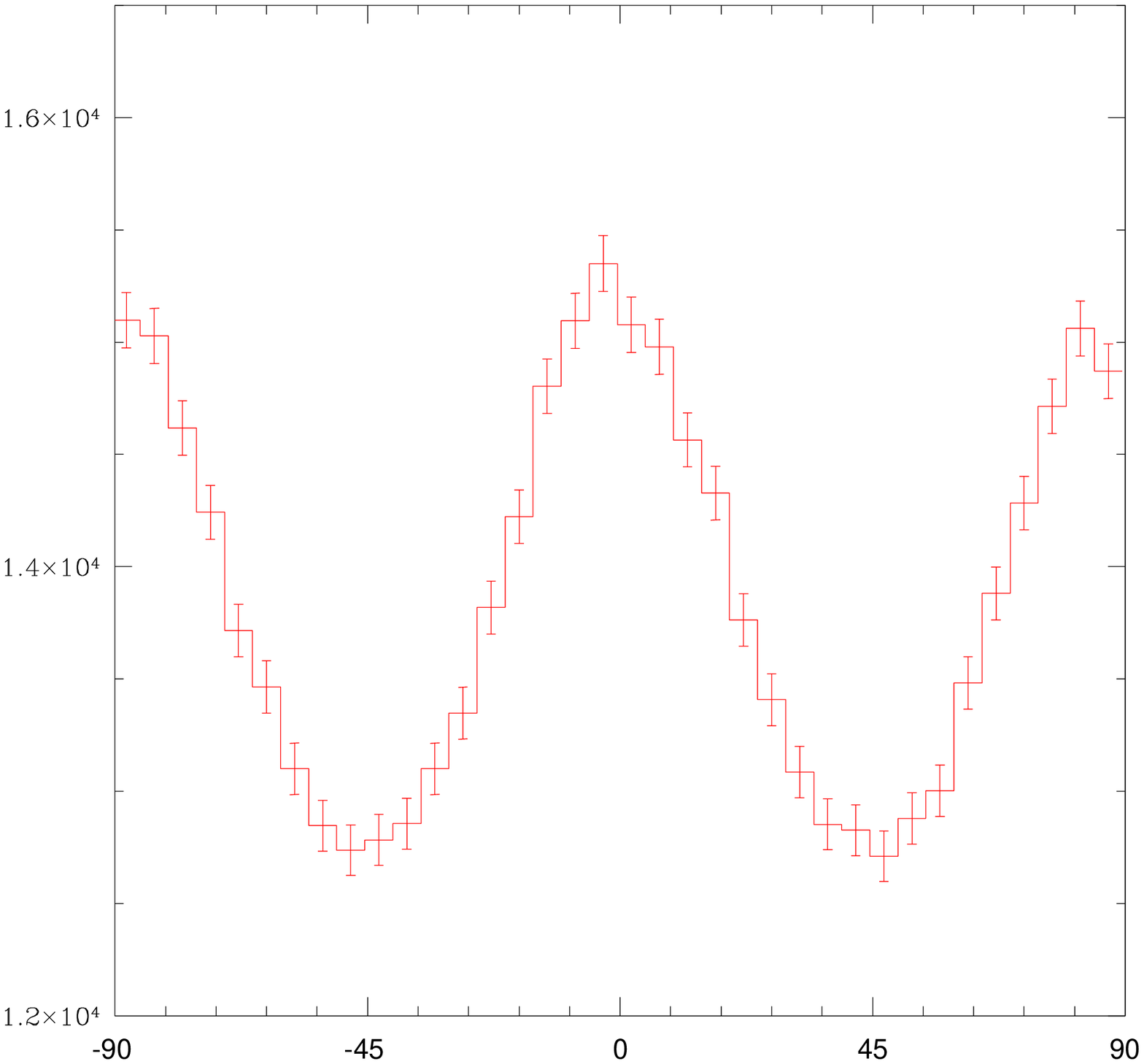,width=9cm,height=9cm}
\end{tabular}
\caption{Histogram of $\alpha$ for a simulated NVSS survey containing 500000 sources with the effects of multiplicative offsets included. (left) $m_Q=0.01$, $m_U=0.0$ and (right) $m_Q=0.1$, $m_U=0.0$. The histogram has a bias which can be modelled by ${\hat N}\propto\cos 4\alpha$. The bias is more visible in the larger case, but the magnitude in the smaller case appears to be similar to that required to be added to the CLEAN bias effect to model the observed data. The errorbars quantify the random Poisson error on each bin.}
\label{fig:multi}
\end{figure}

\begin{figure}
\begin{tabular}{c|c}
\hskip -1.3cm \psfig{figure=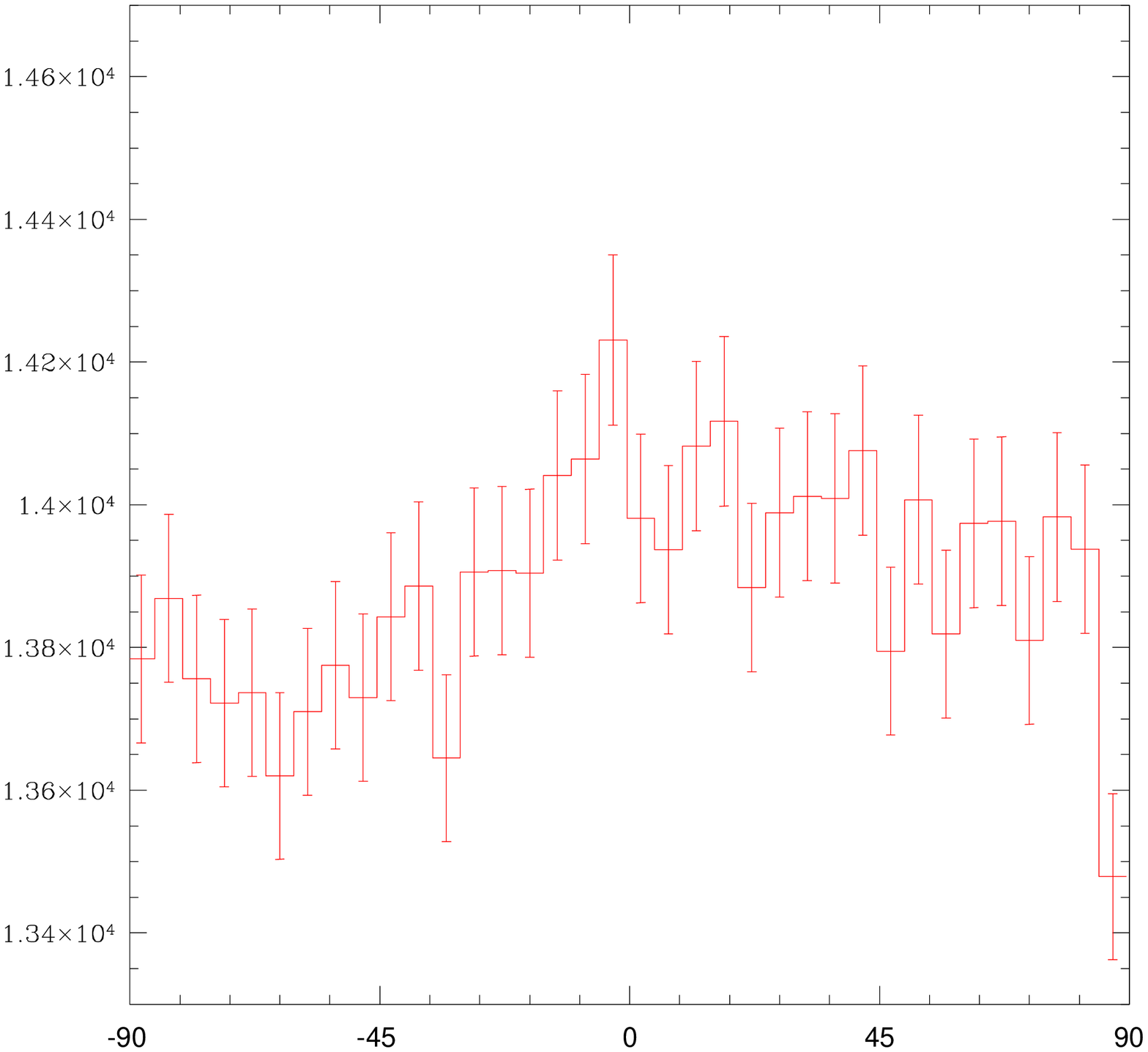,width=9cm,height=9cm}\psfig{figure=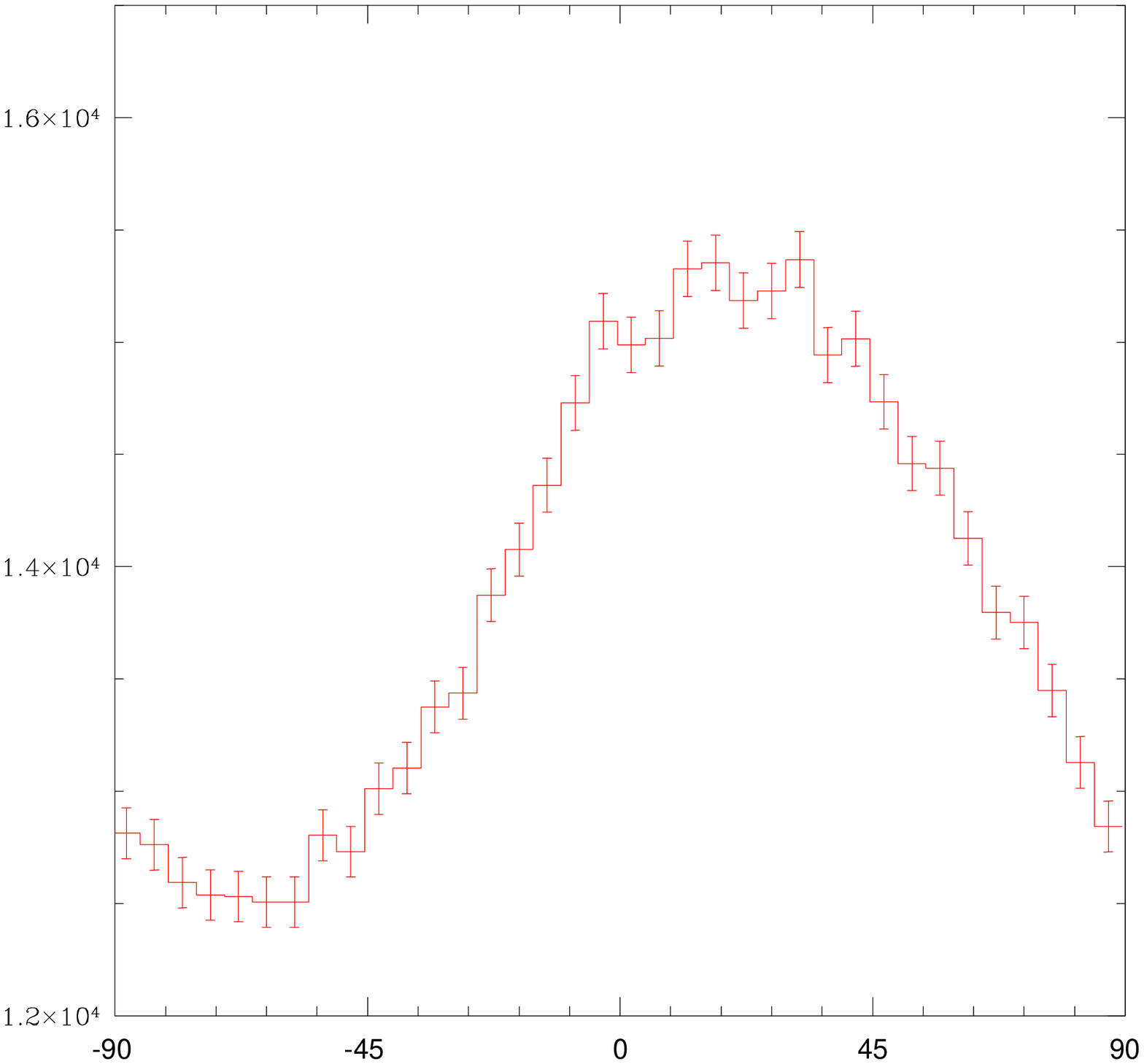,width=9cm,height=9cm}
\end{tabular}
\caption{Histogram of $\alpha$ for a simulated NVSS survey containing 500000 sources with the effects of additive offsets. (left) $a_Q=a_U=5\mu{\rm Jy}$ and (right)  $a_Q=a_U=50\mu{\rm Jy}$.  The histogram has a bias which can be modelled by ${\hat N}\propto\cos 2(\alpha-45^{\circ})$. The bias is much more visible in the larger case, but the magnitude of the smaller case appears to be similar to that required to be added to the CLEAN bias and multiplicative offsets to explain the observed data in the regime where the polarized flux is low. The errorbars quantify the random Poisson error on each bin.}
\label{fig:add}
\end{figure}

In order to confirm our understanding of these effects we have performed the same simulations as discussed in the previous section, but instead of including the effects of CLEAN bias, we have introduced a multiplicative offsets of $m_Q=0.01$ and $m_Q=0.1$, while keeping $m_U$=0.0, as well as additive offsets $a_Q=a_U=5\mu{\rm Jy}$ and $a_Q=a_U=50\mu{\rm Jy}$. The results of these four simulations are presented in Figs.~\ref{fig:multi} and \ref{fig:add}, and these appear to be compatible with the prediction (\ref{alphahist}).

\subsection{Fitting to the observed biases in NVSS}
\label{sec:data}

From the discussion of the preceding sections it appears that a combination of CLEAN bias plus multiplicative and additive offsets can lead to histograms qualitatively similar to those observed in the NVSS data. Therefore, its seems sensible to establish which values of the parameters discussed above best represent the observed biases. We will make an assumption in doing this that there are global, flux independent values for these parameters. This may not be necessarily true since the amount of CLEANing may vary from field to field, and we have not specified the origin of the offsets which could also be different across fields. Moreover, we are presuming that the true values of $\alpha$ are drawn from a uniform distribution.

Using the simulation procedure discussed in section~\ref{sec:clean} we have fitted for the parameters $\epsilon_{\rm C}$, $R=(1+m_U)/(1+m_Q)$, $a_Q$ and $a_U$  in order to give the best fit to the histograms of NVSS polarization position angles. We divided the data into separated histograms: $P<$1mJy, $1-10$mJy, $10-100$mJy and $>$100mJy; each with 36 bins of width $5^{\circ}$. The best fits were obtained for $\epsilon_{\rm C}=14\mu{\rm Jy}$, $R=0.986$, $a_Q=5.1\mu{\rm Jy}$ and $a_U=6.5\mu{\rm Jy}$. This value of $\epsilon_C$ corresponds to $\sim 1$ CLEAN component subtracted per source which seems reasonable at least for the weak sources. Note the extremely small offsets applied; these are many times smaller than the noise level and could not be detected without such a large sample of objects.

\begin{figure}
\psfig{figure=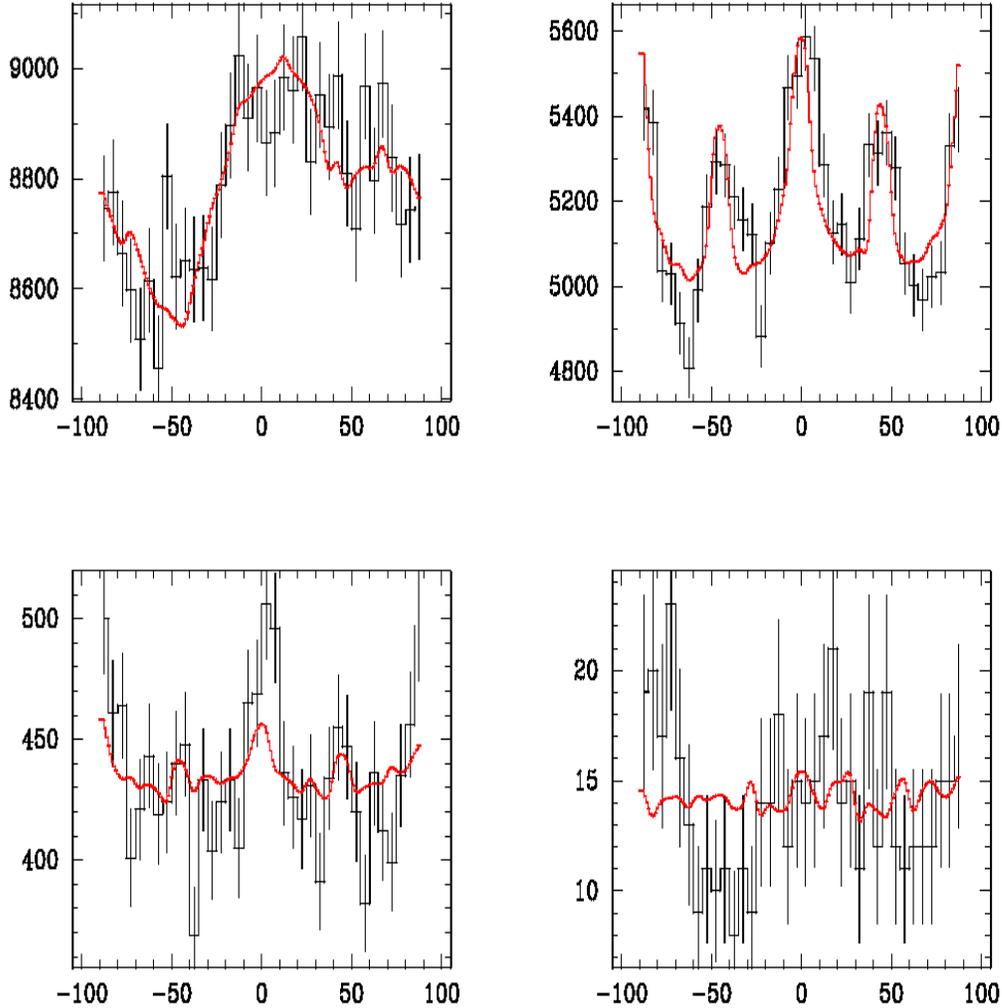,width=16cm,height=16cm,angle=180}
\caption{Four histograms of the NVSS polarization angles split by polarized flux: (top left) $P<1{\rm Jy}$; (top right) $P=1-10{\rm mJy}$; (bottom left) $P=10-100{\rm mJy}$; (bottom right) $P>100{\rm mJy}$. Included also are a single simulation based on the best fitting parameters discussed in the text.}
\label{fig:fit}
\end{figure}

In general, the fit is good; over all four histograms together,
$\chi^2$/dof(degree of freedom)=1.60. However, the two higher flux-density histograms are
less well fit probably since they contribute proportionately less to
the $\chi^2$ as there are fewer sources in the bins. We have
attempted to use a flux-dependent value of $\epsilon_{\rm C}$ to
adjust the level of CLEAN bias and hence improve the fit, but with
limited success. In practice, the situation is likely to be more
complicated than our model in this range because of the added effects
of self-calibration which has been applied during the mapping process
of the high flux-density sources.

\subsection{An attempt at correction of the NVSS polarization data}
\label{sec:corr}

We have attempted to use the fitted parameters deduced in the previous section to correct the NVSS data for the biases in the histogram of $\alpha$. Such a procedure is fraught with danger and should not be taken too seriously since there are two obvious problems: (i) the prescription for including CLEAN bias that we have used is not invertible in the sense that if $|Q_{\rm true}|<\epsilon_{\rm C}$ then $Q_{\rm obs}=0$, and similarly for $U$; (ii) only the ratio $R$ can be directly derived from the observed data and hence any choice of parameters with $m_Q-m_U\approx 0.015$ will lead to the same histogram; (iii) the assumption that the parameters are global and independent of polarized flux is unlikely to be completely true. Nonetheless it is an interesting exercise since an improvement in the properties would suggest that the model fits source by source, rather than just being a pattern spotting exercise on the histograms.

\begin{figure}
\psfig{figure=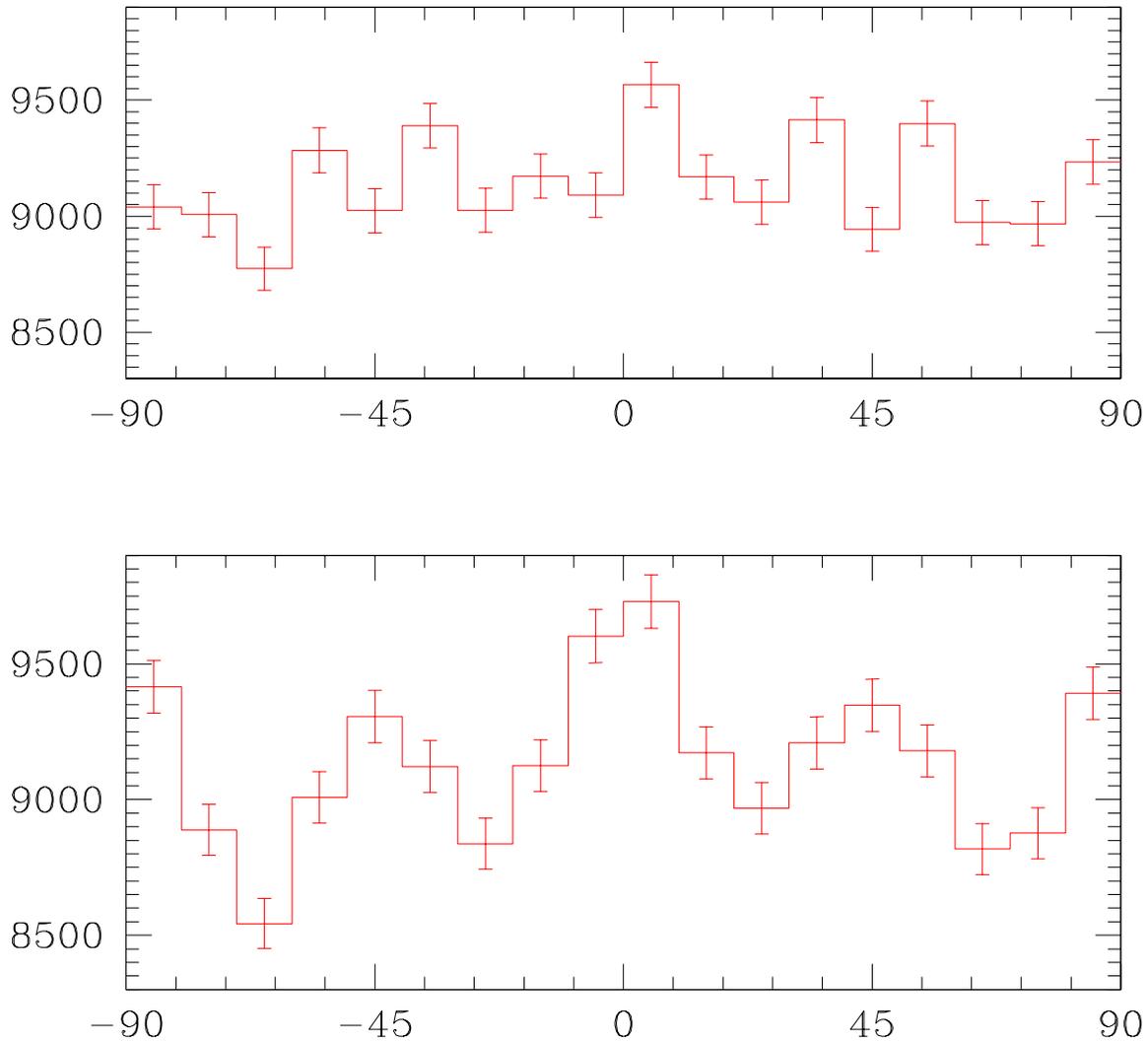,width=16cm,height=16cm}
\caption{Histograms of $\alpha$ before (bottom) and after (top) the ``correction'' in the range $1{\rm mJy}<P<5{\rm mJy}$ for $\epsilon_{\rm C}=14\mu{\rm Jy}$, $m_Q=1.052$, $m_U=1.036$, $a_Q=5.1\mu{\rm Jy}$ and $a_U=6.5\mu{\rm mJy}$.}
\label{fig:corr}
\end{figure}

In order to mitigate against $|Q|,|U|<\epsilon_{\rm C}$, we have chosen to just perform this ``correction'' on sources with $P>1{\rm mJy}$. Moreover, an examination of the two higher flux histograms in Fig.~\ref{fig:fit} suggests that if one  also constrains $P<5{\rm mJy}$, one might expect to be at least partially successful. We have chosen $m_Q=1.052$ and $m_U=1.036$ which satisfy the above constraint. The resultant histogram is presented in Fig.~\ref{fig:corr}; before the correction the $\chi^2/{\rm dof}$ is  $\approx 9$ and afterwards it is $\approx 4$ which suggests that, although the procedure we have used is obviously not the complete story, there is some reason to believe that it is along the right lines. 

We note that although the improvement in the $\chi^2/{\rm dof}$ is more than a factor of two, the change to each of the individual polarization angles is very small. We find that the $\langle(\Delta\alpha)^2\rangle^{1/2}\approx 0.3^{\circ}$ with very few angles changing by more than $2^{\circ}$. It quite interesting to see that such small systematic changes in the values of $\alpha$ can lead to such a large effect on the $\chi^2/{\rm dof}$.

\section{Discussion and conclusions}

There is a striking systematic in the histogram of $\alpha$ for measurements presented in the NVSS catalogue. We have argued that this is as result of a number of very small effects that are substantially below the noise level for each source, but which lead to observable effects of high significance when ``summed'' up over the whole catalogue. In attempting to make a correction for the effect we find that very small changes in the value $\alpha$ can lead to a more uniform histogram. We suggest that CLEAN bias is the chief systematic effect since it preferentially selects sources with $\alpha=-90^{\circ}$, $-45^{\circ}$, $0^{\circ}$, $45^{\circ}$ and $90^{\circ}$. Rather than making the correction discussed in section~\ref{sec:corr} which we have already pointed out has a number of flaws when considered source-by-source, the effects of CLEAN bias can be avoided by extracting the sources in the visibility plane. This was how the final values of $Q$ and $U$ were extracted for the JVAS/CLASS survey (Jackson et al 2007) during the analysis of which we first became aware of such an effect, albeit for a much smaller number of sources. Adopting a visibility plane approach for the NVSS would be much more difficult since there are typically a large number of sources in each field, some with complex structure, which was not the case in JVAS/CLASS. We emphasize that the effects which we are discussing are generally $<<2^{\circ}$ for any given source and that using the NVSS catalogue for normal astrophysical purposes is unlikely to lead to a substantial error unless the whole catalogue is being used statistically to reduce the random errors.

We note that the kind of biases which we have highlighted may serve as a lesson for the future. It appears that before making strong claims on the basis of statistical analysis of polarization, data should first be carefully assessed for the kind of effects discussed here. This is particularly relevant since new instruments are being designed to perform observations of very large numbers of  polarized radio sources to study magnetic fields, and of the cosmic microwave background (CMB) to search for gravitational waves via the B-mode signatures. Both these will require exquisite control of systematics.

\begin{figure}
\begin{tabular}{c|c}
\hskip -1.3cm \psfig{figure=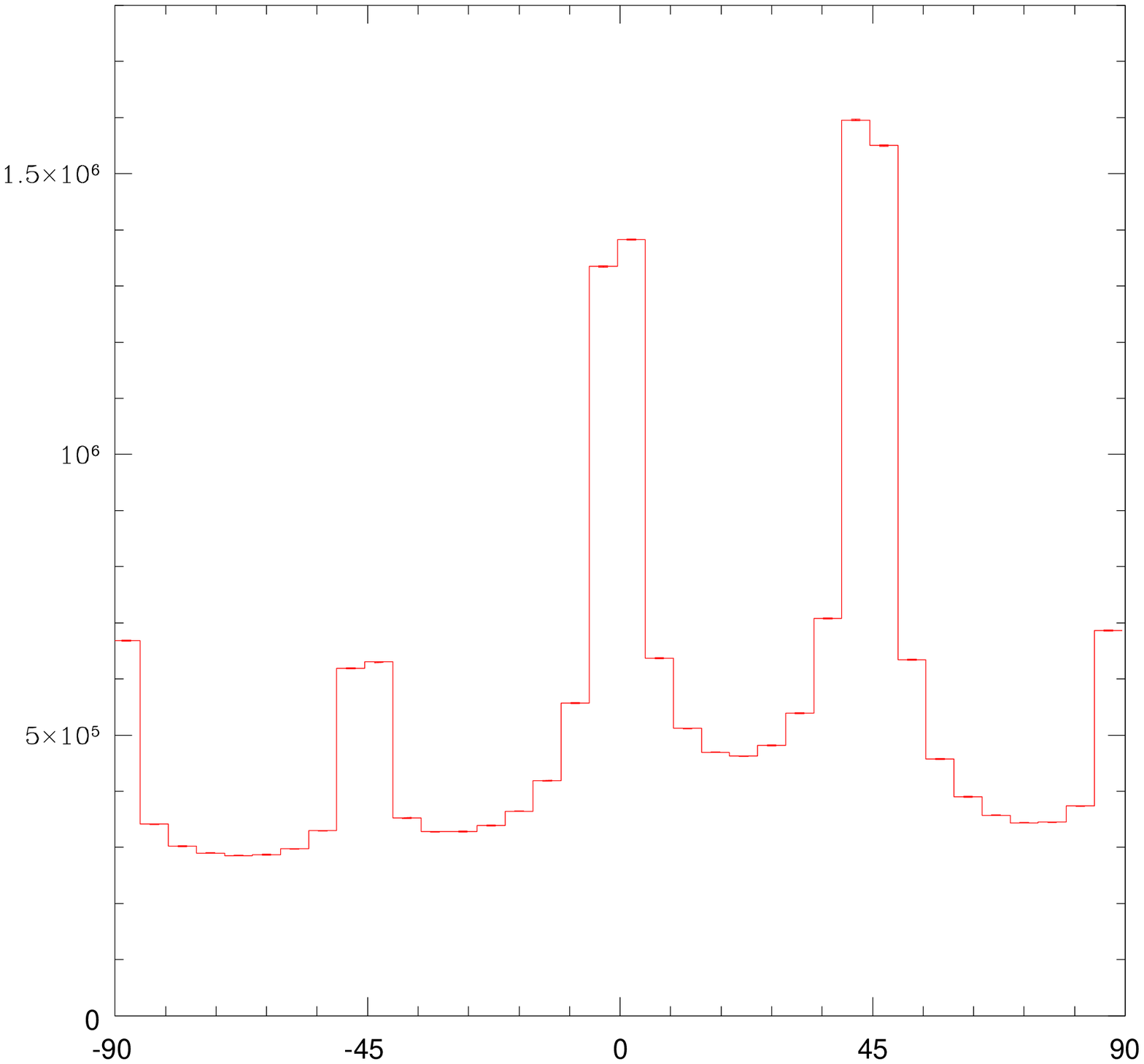,width=9cm,height=9cm}\psfig{figure=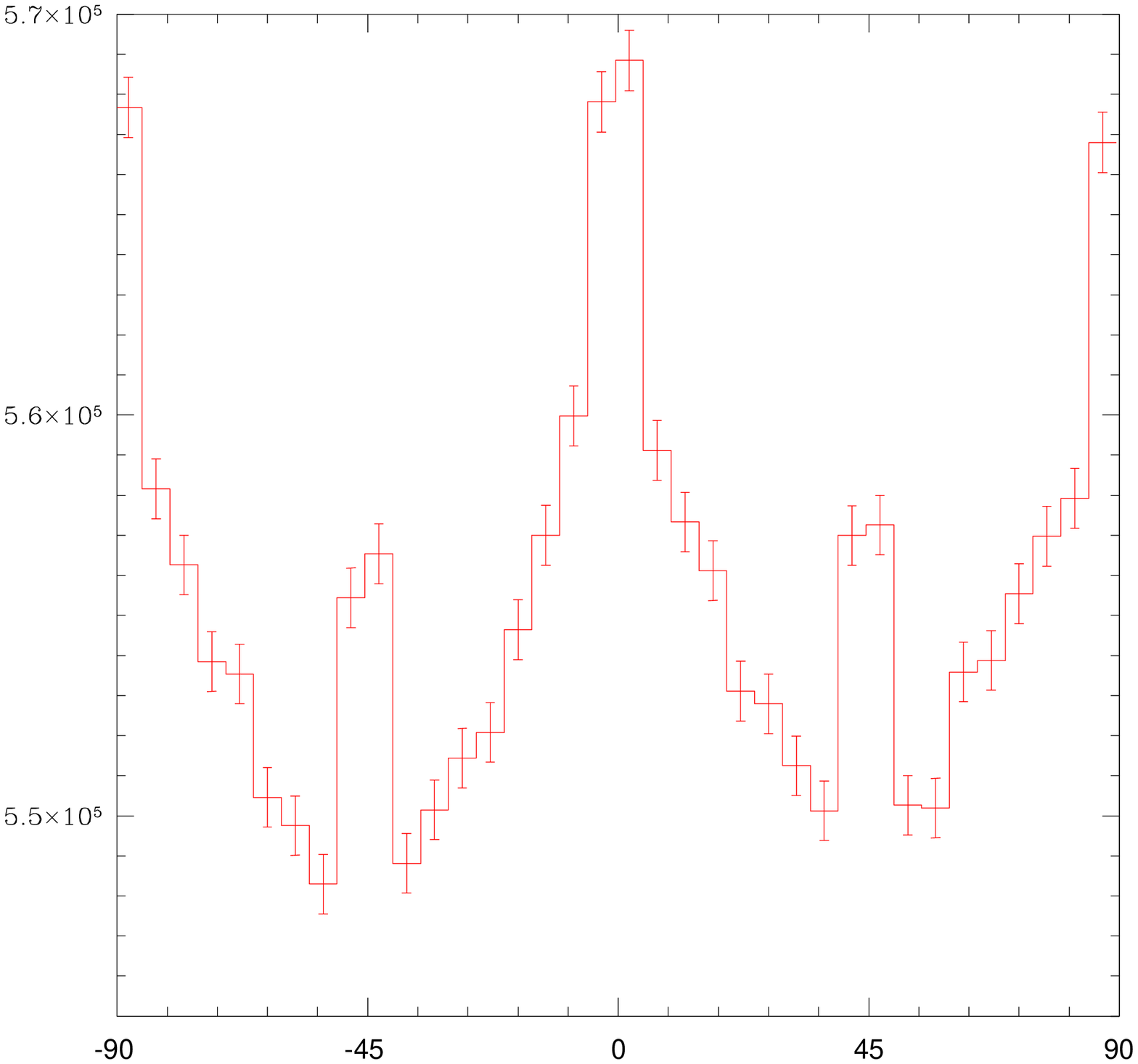,width=9cm,height=9cm}
\end{tabular}
\caption{Histogram of $\alpha$ for a simulated SKA survey containing $2\times 10^7$ sources with systematics included. (left) $\epsilon_{\rm C}=15\mu{\rm Jy}$, $m_Q=0.01$, $m_U=0.0$, $a_{Q}=5\mu{\rm Jy}$ and $a_{U}=6.5\mu{\rm Jy}$ as was found to represent those found in the NVSS survey (right) $\epsilon_{\rm C}=50{\rm nJy}$, $m_Q=0.01$, $m_U=0.0$, $a_{Q}=17{\rm nJy}$ and $a_{U}=22{\rm nJy}$ which correspond to a scaling of the noise level expected for the SKA survey, relative to the NVSS survey.}
\label{fig:sim}
\end{figure}

As a final point we discuss how this kind of systematic might affect future high precision observations of polarization. Let us first consider a point source survey with the same systematic biases (that is, the same values of $\epsilon_{\rm C}$, $m_{Q}$, $m_{U}$, $a_{Q}$ and $a_{U}$) as in the NVSS, but with $2\times 10^7$ sources between $P_{\rm min}=10\mu{\rm Jy}$ and $P_{\rm max}=100{\rm mJy}$ with a noise level of $\sigma_{Q}=\sigma_{U}=1\mu{\rm Jy}$ such as might be relevant to the SKA. In such a survey, which is compatible with the distribution of the NVSS observed at higher flux densities, the polarization would be detected at $S/N>10$ for all sources. It has been shown that this would allow one to constrain the rotation measure (RM) to $\pm 5\,{\rm rad}\,{\rm m}^{-2}$ using a survey at $\lambda=21{\rm cm}$ and $\Delta\lambda/\lambda=0.25$ (Beck and Gaensler, 2004). A simulated histogram of position angles is presented in the left hand column of Fig.~\ref{fig:sim} showing extremely strong biases, with peaks at $0^{\circ}$ and $45^{\circ}$ which are a factor $\sim 3$ higher than the rest of the histogram. Such an effect would clearly be catastrophic for the extraction of science from such a survey, but is probably unrealistic since the levels of CLEAN bias and the additive offsets are likely to be defined by the noise level, or possibly the flux of the source under consideration. We have also included in the right hand column of Fig.~\ref{fig:sim} the results of a simulation with the values of $\epsilon_{\rm C}$, $a_{Q}$ and $a_{U}$ scaled down by a factor 300 which is the ratio of the noise level in the proposed SKA survey to that in the NVSS. The amplitude of bias in this case is of similar order of magnitude to that found in the NVSS although the structure of the peaks in the histogram is much better defined due to the increased number of sources in each bin.

It is interesting to speculate on how the smaller of these two biases might affect the extraction of science from such a survey. If the bias is similar in character to that in the NVSS then $\langle(\Delta\alpha)^2\rangle\approx 0.3^{\circ}$, that is, there is a offset, dependent on the intrinsic position angle with this RMS. This will lead to an offset in the measured RM and its error. The noise error on each position angle will be around $5^{\circ}$ corresponding to the the $S/N>10$ which is much larger than the offset/increase in the noise. However, it is planned for the sets of rotation measures to be used together to reduce the random errors and deduce information about magnetic fields. If $\sim 300$ sources were added together then the resultant random errors would be comparable to the offset.

The other area where high precision measurements of polarization are planned is the CMB. There position angle accuracies $<<0.3^{\circ}$ will be required to probe primodial gravitational waves corresponding to scalar-to-tensor ratios, $r\sim 10^{-3}$ which might be possible within planned polarization satellites. In this application it is unlikely that CLEAN bias will be a problem since power spectrum estimates will be made directly from the ``dirty'' images and hence do not use the CLEAN algorithm, but the multiplicative and additive offsets are likely to be an issue. We plan to investigate whether the kind of techniques used here can be used to search for subtle systematic effects in the measured polarization.

\section*{Acknowledgements}

We would like to thank Bill Cotton for helpful comments, and Peter Wilkinson for providing considerable encouragement.

\section*{References}

\noindent Beck, R., 2001, Space Sci. Rev. 99, 243

\noindent Beck, R., Gaenseler, B.M., 2004, New Ast. Rev. 48, 1289 

\noindent Birch, P., 1982, Nature 298, 451

\noindent Carilli, C., Rawlings, S., 2004,''Science with the SKA'', New Ast. Rev. 48, 11-12

\noindent Carilli C., Taylor, G.B., 2002, ARA\&A 40 319

\noindent Condon, J.J., Cotton, W.D., Greisen, E.W., Yin, Q.F., Perley, R.A., Taylor, G.B., Broderick, J.J., 1998, AJ 115, 1693

\noindent Feretti, L., Burigana, C., Ensslin, T.A., 2004, New Ast. Rev. 48, 1137

\noindent Feretti, L., Johnston-Hollitt, M., 2004, New Ast. Rev. 48, 1145

\noindent Hogbom, J.A., 1974, A\&AS 15. 417

\noindent Hutsemekers, D., Cabanac, R., Lamy, H., Sluse, D., 2005, A\&A 441, 915

\noindent Jackson, N., Battye, R.A., Browne, I.W.A., Joshi, S.A., Muxlow, T.W.B., Wilkinson, P.N., 2007, MNRAS 376, 371

\noindent Joshi, S.A., Battye, R.A., Browne, I.W.A., Jackson, N.,  Muxlow, T.W.B., Wilkinson, P.N., 2007, MNRAS 380, 162

\noindent Kendall, D., Young., G., 1984, MNRAS 207, 63

\noindent Phinney, E., Webster, R., 1983, Nature 201, 735

\noindent White, R.L., Becker R.H., Helfand, D.J., Gregg, M.D., 1997, ApJ 475, 479

\bigskip
\end{document}